\documentclass[lettersize,journal]{IEEEtran}

\usepackage{amsmath,amsfonts}

\usepackage{array}
\usepackage{textcomp}
\usepackage{stfloats}
\usepackage{url}
\usepackage{verbatim}
\usepackage{cite}

\usepackage{makecell}
\usepackage{float}
\usepackage{multicol}
\usepackage{amsmath,amssymb}
\usepackage{subfigure}
\usepackage{stfloats}
\usepackage{flushend}

\usepackage{balance}

\usepackage{amsthm}
\usepackage{makecell}
\usepackage{float}
\usepackage{graphicx,times,amsmath,booktabs}
\usepackage{multicol}

\usepackage{epstopdf}
\usepackage{stfloats}

\usepackage{balance}
\usepackage{xcolor}

\usepackage[ruled,linesnumbered]{algorithm2e}
\usepackage{amsmath}

\usepackage{cite}

\usepackage{amsmath}
\usepackage{amsthm}
\usepackage{makecell}
\usepackage{float}
\usepackage{graphicx,times,amsmath,booktabs}
\usepackage{multicol}
\usepackage{amsmath,amssymb}
\usepackage{subfigure}
\usepackage{stfloats}
\usepackage{flushend}
\usepackage{epstopdf}
\usepackage{balance}
\usepackage{xcolor}

\newtheorem{theorem}{Theorem}
\newtheorem{lemma}{Lemma}

 \usepackage[ruled,linesnumbered]{algorithm2e} 
\usepackage{amsmath}

\usepackage{cite}

\renewcommand{\algorithmcfname}{Algorithm}
\def\BibTeX{{\rm B\kern-.05em{\sc i\kern-.025em b}\kern-.08em
		T\kern-.1667em\lower.7ex\hbox{E}\kern-.125emX}}

\begin{document}

\title{Performance Analysis of Uplink/Downlink Decoupled Access in Cellular-V2X Networks}
%
%

\author{Luofang Jiao,~\IEEEmembership{Member,~IEEE,} Kai Yu,~\IEEEmembership{Member,~IEEE,} Jiacheng Chen,~\IEEEmembership{Member,~IEEE,} \\Tingting Liu,~\IEEEmembership{Member,~IEEE,} Haibo Zhou,~\IEEEmembership{Senior Member,~IEEE,} and Lin Cai,~\IEEEmembership{Fellow,~IEEE}
\IEEEcompsocitemizethanks{
\IEEEcompsocthanksitem	L. Jiao, K. Yu, and H. Zhou (Corresponding author) are with the School of Electronic
Science and Engineering, Nanjing University, Nanjing 210023, China.
E-mail: \{luofang\_jiao, kaiyu\}@smail.nju.edu.cn, haibozhou@nju.edu.cn.

\IEEEcompsocthanksitem T. Liu is with School of Electronic Science and Engineering, Nanjing University, Nanjing 210023, CHINA, and is also with School of Computer Science, Nanjing University of Posts and Telecommunications, Nanjing 210023, China. E-mail: liutt@njupt.edu.cn.

\IEEEcompsocthanksitem   J. Chen is with the Peng Cheng Laboratory, Shenzhen 518000, China.
E-mail: chenjch02@pcl.ac.cn.

\IEEEcompsocthanksitem L. Cai is with the Department of Electrical and Computer Engineering, University of Victoria, Victoria, BC V8P 5C2, Canada. E-mail: cai@ece.uvic.ca.

}}

\IEEEtitleabstractindextext{%
\begin{abstract}
This paper firstly develops an analytical framework to investigate the performance of uplink (UL) / downlink (DL) decoupled access in cellular vehicle-to-everything (C-V2X) networks, in which a vehicle's UL/DL can be connected to different macro/small base stations (MBSs/SBSs) separately. Using the stochastic geometry analytical tool, the UL/DL decoupled access C-V2X is modeled as a Cox process, and we obtain the following theoretical results, i.e., 1) the probability of different UL/DL joint association cases i.e., both the UL and DL are associated with the different MBSs or SBSs, or they are associated with different types of BSs; 2) the distance distribution of a vehicle to its serving BSs in each case; 3) the spectral efficiency of UL/DL in each case; and 4) the UL/DL coverage probability of MBS/SBS. The analyses reveal the insights and performance gain of UL/DL decoupled access. Through extensive simulations, \textcolor{black}{the accuracy of the proposed analytical framework is validated.} Both the analytical and simulation results show that UL/DL decoupled access can improve spectral efficiency. The theoretical results can be directly used for estimating the statistical performance of a UL/DL decoupled access C-V2X network.
\end{abstract}

}

\maketitle

\IEEEdisplaynontitleabstractindextext

\IEEEpeerreviewmaketitle

\vspace{0.8cm}
\IEEEraisesectionheading{\section{Introduction}\label{sec:introduction}}

\subsection{Background and Motivation}
\IEEEPARstart C{ellular} vehicle-to-everything (C-V2X) is vital for autonomous driving and intelligent transportation systems (ITS) in the fifth-generation mobile networks (5G) and beyond \cite{mollah2020blockchain, kim2019new, chen2020vision, zhao2023fully}. 
\textcolor{black}{To satisfy the stringent quality-of-service (QoS) requirements of various C-V2X applications \cite{kang2019toward, zhang2019stochastic}, small base stations (SBS) can be deployed alongside the macro base stations (MBS) \cite{bhushan2014network}.}
Decoupled access of uplink (UL) and downlink (DL) is a promising technology and its performance has been investigated in typical cellular networks \cite{boccardi2016decouple, jiao2022spectral}.
In decoupled access networks, a user equipment (UE) is allowed to simultaneously connect with an MBS and an SBS for separate UL/DL transmission \cite{elshaer2014downlink}, so both UL and DL can utilize the most appropriate BSs.

Given the increased spectrum frequency and the heightened demands of vehicular users' services, the C-V2X is gradually evolving into a more complicated heterogeneous network including MBSs and SBSs \cite{boccardi2016decouple}, \cite{yan2017novel}, \cite{zhou2020evolutionary}. The traditional DL receiving power-based UL association mode is already not appropriate for C-V2X.
In this paper, we study the UL/DL decoupled access C-V2X networks to provide an in-depth understanding of decoupled access in C-V2X through theoretical analysis and to investigate its potential performance gain in terms of spectral efficiency (SE).
The major challenge comes from the complexity of distributions of roads, vehicles, SBSs and MBSs \cite{zhou2020evolutionary}, \cite{he2015delay}.
\textcolor{black}{
Specifically, in C-V2X networks, vehicles and SBSs are randomly distributed along the roads following a Poisson point process (PPP). \textcolor{black}{The roads can be modeled as lines of a Poisson line process (PLP), and the roads and the MBSs are randomly distributed on a two-dimensional (2D) plane following a PLP and PPP, respectively \cite{dhillon2020poisson}.} In contrast, in typical cellular networks, both UEs and BSs are assumed independently distributed on a 2D plane following a PPP.}
Therefore, the traditional UL/DL decoupled analysis methods in cellular networks are not applicable in C-V2X.

\vspace{-0.3cm}
\subsection{Related Work}
\textcolor{black}{Recently, the UL/DL decoupled access has been shown to outperform traditional coupled access in cellular networks.
For example, Zhang \emph{et al.} theoretically investigated the UL performance improvement under the UL/DL decoupled mode over the coupled access mode in \cite{zhang2017uplink}.
Li \emph{et al.} studied the UL SE in multiuser multiple-input multiple-output (MIMO) heterogeneous networks and revealed the superiority of decoupled mode over coupled mode \cite{li2018uplink}.
In \cite{smiljkovikj2015analysis}, Smiljkovikj \emph{et al.} firstly derived the probability of different cases of UL/DL decoupled association in heterogeneous wireless networks.
A more recent work \cite{sattar2019spectral} presented an in-depth analysis of SE in UL/DL decoupled cellular networks.
In \cite{yu2019fully}, Yu \emph{et al.} proposed a novel fully-decoupled RAN (FD-RAN) architecture for 6G. 
\textcolor{black}{Shi \emph{et al.} proposed a decoupled access scheme with enhanced energy efficiency for cellular-enabled UAV communication networks \cite{shi2021energy}.}
Different from the works mentioned above, our analysis jointly considers UL and DL in the decoupled access C-V2X networks, in which distributions of roads, vehicles, SBSs, and MBSs are different types of Poisson processes and are independent.}

C-V2X is a special application scenario of cellular networks with UEs, i.e., vehicles, moving fast along the roads.
In \cite{9709535}, Pan \emph{et al.} presented the coverage probability analysis for safety messages-prioritized C-V2X communications.
In \cite{chetlur2018coverage} and \cite{chetlur2019coverage}, Chetlur \emph{et al.} presented downlink coverage and rate analysis in C-V2X.
Sial \textit{et al}. introduced a tractable analytical framework for performance of C-V2X networks over shared V2V and cellular uplink channels \cite{8859331}.
In \cite{9348103}, Liu \emph{et al.} studied millimeter wave (mmWave) communications of uplink by leveraging the stochastic geometry theory in C-V2X.
Yu \textit{et al}. introduced a reinforcement learning-based RAN slicing framework for V2X communications with the aid of UL/DL decoupled access \cite{9599513}.
It is demonstrated that this technology can significantly improve load balancing and reduce the transmit power.
Despite the contributions of existing works on C-V2X, there still lacks a thorough performance analysis for the UL/DL decoupled access C-V2X networks.

\vspace{-0.35cm}
\subsection{Contributions}

This paper fills the gap in the performance analysis of decoupled access in C-V2X networks.
First, we use variants of the Poisson process to model the distributions of roads, vehicles, SBSs, and MBSs in the C-V2X.
\textcolor{black}{Note that vehicles and SBSs are actually modeled by the `doubly stochastic' Cox process since they are distributed only alongside the roads.
Then, stochastic geometry is adopted as the mathematical tool to obtain the following four key results,} namely 1) the probabilities of four different UL/DL joint association cases (i.e., UL=DL=MBS/SBS, UL=MBS/SBS, and DL=SBS/MBS), given by Lemma 1-3; 2) distribution of vehicle's distance to its serving BSs, given by Lemma 4-6; 3) SE of UL/DL in all cases, given by Theorem 1-6; and 4) coverage probability (CP) of UL/DL of MBS/SBS, given by Theorem 7-10.
Besides, SE of coupled access associated with MBS/SBS is given by Corollary 1 such that the performance of decoupled and coupled access can be compared in theory under the same analytical framework.
To summarize, the contributions of this paper are listed as follows:

\begin{itemize}
	
	\item \textcolor{black}{We propose a tractable analytical framework for UL/DL decoupled access in C-V2X considering random distributions of roads, vehicles, SBSs, and MBSs.}

	\item We provide theoretical results on UL/DL decoupled joint association probability, distance distributions for all decoupled cases, SE of all links and CP are given. The effect of speed is analyzed to study the mobility of vehicular networks. The theoretical results can be directly used for estimating the statistical performance of a UL/DL decoupled access C-V2X network.
	
	\item We conduct extensive simulations to verify the accuracy of the proposed analytical framework, and the results show better performance in UL/DL decoupled C-V2X networks in terms of load balance and SE, which will shed light on the further study of UL/DL decoupled access in C-V2X.
\end{itemize}

The remainder of this paper is organized as follows. Section 2 introduces the network model. In Section 3, the analysis of decoupled access in C-V2X is presented. The simulation results are given and discussed in Section 4. The paper is concluded in Section 5.

\begin{figure}[t]
	\centering
	\centerline{\includegraphics[width=0.8\hsize]{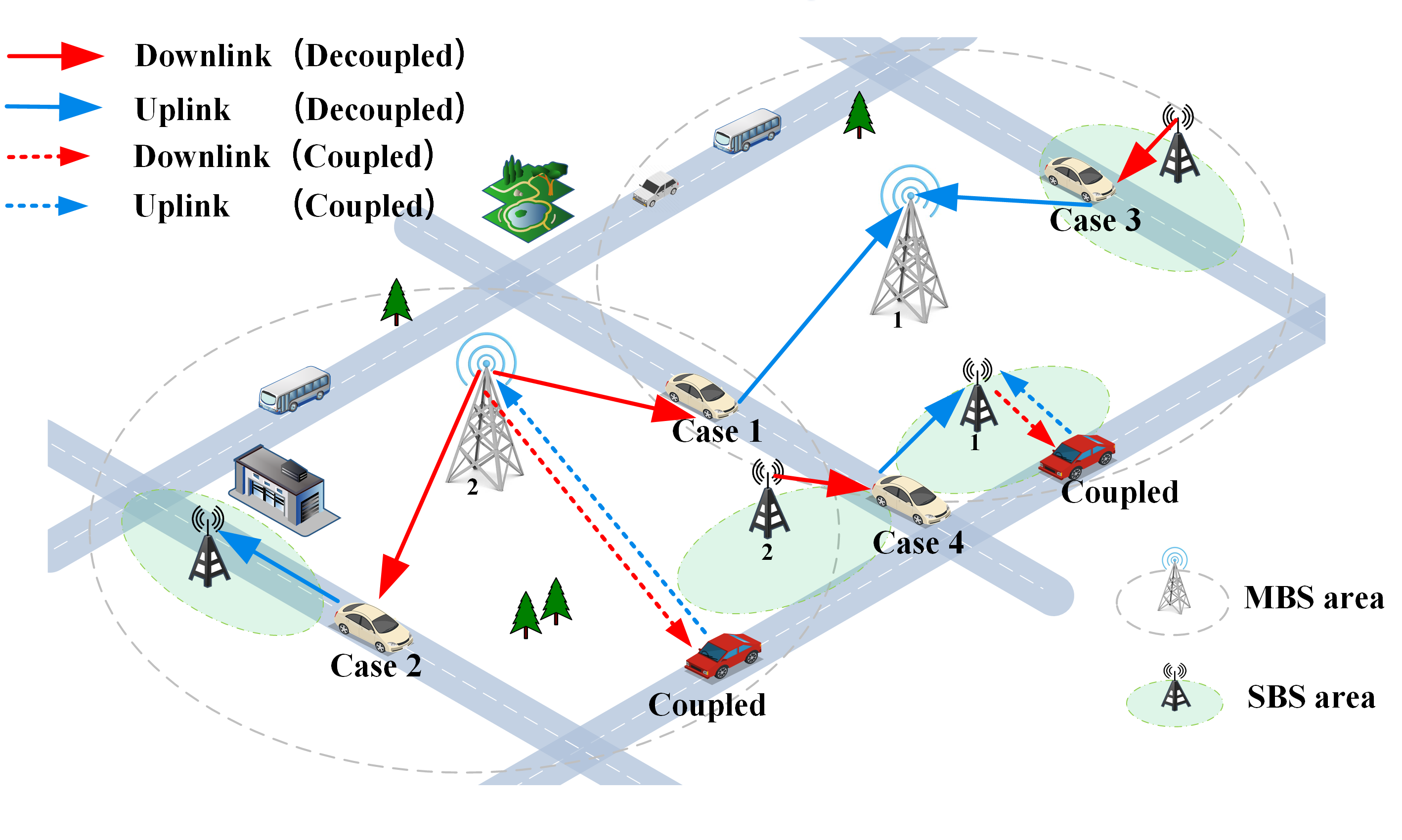}}
	\caption{Illustration of UL/DL decoupled access in C-V2X networks.}
	\label{system-model}
\end{figure}

\begin{table}[h] 
	\centering 

	\color{black}\caption{A LIST OF MAJOR SYMBOLS}
	\begin{tabular}{l|l }
		\toprule 	
		\hline \label{MAJOR SYMBOLS}
		Notation & Description\\
		\hline
		$ v$, $\lambda$& The speed of vehicle. The density of vehicle and BS. \\
		$ x^{*} $ & The location of closest BS to vehicle. \\
		$ P $ & Received and transmit power of vehicle, SBS, and MBS. \\
		$ G $, $ g $ & The main lobe gain and side lobe gain.\\ 
		$ H $, ${\chi} $ & The channel gain and Nakagami-m fading gain.\\
		$ \Gamma\left ( \cdot \right ) $ & The gamma distribution function of a random variable.\\
	    $ t $ & The predetermined threshold $ t $ of CP or SE. \\
		$  \Xi _{{l}} $  & The Poisson line process of lines.\\
		$\Xi_{l_{i}}^{S}$, $  \Xi_{l_{i}}^{V} $ & The Poisson point processes of vehicles and SBSs on line $ i $.\\
		$ \Theta_{V}$, $ \Theta_{S}$ & The Cox processes of vehicles and SBSs. \\
		$ I_i $ & Interference from set $ i $.\\
	    $ Pr\left ( \cdot  \right ) $ & The probability of a random variable.\\
	    $ E\left ( \cdot  \right ) $ & The expectation of a random variable.\\
	    $ F\left ( \cdot  \right ) $ & The cumulative distribution function of a random variable.\\
	    $ f\left ( \cdot  \right ) $ & The probability density function of a random variable.\\
	    $ \zeta _{I}\left( \cdot  \right) $ & Laplace transform of interference $ I $.\\
	    $ \mathbb{K}\left ( \cdot;\cdot;\cdot;\cdot \right ) $ & The newly defined function for $ \zeta _{I}\left( \cdot  \right) $. \\
	    $ \tau _{i}^n $ & The SE of n, $ n \in \{DL, UL\} $ for Case $ i $.\\
		\hline
		\bottomrule
	\end{tabular}
\vspace{-0.7cm}
\end{table}

\vspace{-0.2cm}
\section{System Model}

\subsection{Modeling of C-V2X Network}
We consider a two-tier heterogeneous UL/DL decoupled access C-V2X network, which consists of MBSs, SBSs, and vehicles. For the convenience of readers, the mathematical parameters used in this paper are summarized in Table \ref{MAJOR SYMBOLS}. The MBSs are randomly distributed on a 2D plane. We first model the spatial layout of MBSs by a 2D PPP $\Phi_M$  with density $\lambda_m$ in the Euclidean plane. The locus of points that are geometrically closer to the selected MBSs than to any other MBS form a Voronoi cell. Then we model the roads by a PLP $  \Xi _{{l}} $ with intensity $\lambda_l$ \cite{chetlur2018characterization}. Since the SBSs and vehicles are distributed along the roads, we model the locations of vehicles and SBSs on each line $L_{i}\in \Xi _{l} $ by 1D PPPs $\Xi_{l_{i}}^{S}$ and $  \Xi_{l_{i}}^{V} $ with densities $\lambda_s$ and $\lambda_v$, respectively. \textcolor{black}{Both $\Xi_{l_{i}}^{S}$ and $  \Xi_{l_{i}}^{V} $ follow 1D PPPs with constant densities. Therefore, the locations of vehicles and SBSs follow Cox processes $ \Theta_{V}= \{\Xi_{l_{i}}^{V}\}_{\{l_i\in \Xi _{l}\}} $, $ \Theta_{S}= \{\Xi_{l_{i}}^{S}\}_{\{l_i\in \Xi _{l}\}}  $, respectively \cite{dhillon2020poisson}. It is worth mentioning that the Cox processes  $ \Theta_{V}$ and $ \Theta_{S}$ are stationary and this property will be used in \textcolor{black} {calculating the intensity in Section 2.2} \cite{morlot2012population}}. In C-V2X, the mobility of vehicles can affect the macroscopic traffic density \cite{tang2020deep}.

For our study, we choose one vehicle as the typical node and translate it to the origin $ o\equiv \left ( 0,0 \right ) $. Since the typical vehicle must be located on a road, and the road must pass through the origin, we set this road as a typical road $l_{o}$. The roads that do not pass through the origin are referred to as the other roads. Under Palm probability of vehicle point process, the vehicles and SBSs on the typical road form the point process $ \Theta_{V_o}=\Theta_{V}\cup \Xi_{L_{o}}^{V} $ and $ \Theta_{S_o}=\Theta_{S}\cup \Xi_{L_{o}}^{S}$. \textcolor{black}{Both the processes $ \Theta_{V_o}$ and $ \Theta_{S_o}$ are the superposition of a Cox process $ \Theta_{i} $ and an independent 1D PPP $ \Xi_{L_{o}}^{i}$ on line $ L_{o} $, where $ i\in \{V,S\}  $.} \textcolor{black}{The results are derived via the application of Slivnyak's theorem to the line process $  \Xi _{{l}} $ \cite{chetlur2018coverage}, \cite{morlot2012population}.}

\vspace{-0.2cm}

\subsection{Propagation Model}

We assume that the MBSs transmit to vehicles on the typical road and the other roads with the same power. Since SBSs and vehicles mainly communicate with the nodes on the road where they are located, we assume that SBSs and vehicles use transmit beamforming to maximize the received power at the receivers. \textcolor{black}{Due to the non-identical behavior of the individual transmit and receive chains of SBSs and vehicles, we assume non-reciprocity channels.} 

We assume that all the MBSs have the same transmitting power $ {P_M} $. All the SBSs and the vehicles have the same transmit power $ {P_S} $ and $ {P_V} $, respectively. \textcolor{black}{Because the SBSs and vehicles use directional antennas \cite{ou2019gps}, we assume a beam pattern whose main lobes are along the roads on which the nodes are located and the side lobes facing other directions.} We use $ G_{S,0}$, $ G_{V,0} $ and $ G_{S,1}$, $ G_{V,1} $ to denote the main lobe gains and side lobe gains of SBSs and vehicles, respectively. Since the MBSs have a larger service range and are randomly arranged, we assume that the MBSs have omnidirectional antennas and use $ G_M$ to denote the gains of MBSs.
The general Nakagami-m fading is chosen to model the wide-range fading environment \cite{beaulieu2005efficient}. We use $ m_M $ to denote the fading parameter for the links between the typical vehicle and MBSs (VM) and $ H_M $ to denote the channel fading gains for DL of VM. The SBSs on the typical road are more likely to have line-of-sight (LOS) links to the typical vehicle than to vehicles on the other roads. Therefore, we denote the UL/DL fading parameters between the SBSs and vehicles as $ m_{S,0} $, $ m_{V,0} $ on the typical road, and $ m_{S,1} $  $ m_{V,1} $ on the other roads. \textcolor{black}{
The $ H_{S,0} $, $ H_{S,1} $ denote the corresponding fading gains in DL for the links between the typical vehicle and SBSs on the typical road (VST), and  the links
between the typical vehicle and SBSs on the other roads (VSO), respectively. The $ H_{V,0} $ and $ H_{V,1} $ denote the corresponding fading gains in UL for the links of VST and VSO, respectively.} The fading gains follow a Gamma distribution and its probability density function (PDF) is \cite{duong2012cognitive}
\textcolor{black}{\begin{equation}
f_{H_i}=\frac{m_{i}^{m_i}h^{m_i-1}}{\Gamma\left ( m_i \right ) }e^{-m_i h}, i \in \left \{ {M,S,V} \right \}.
\end{equation}}
For the wireless links, we consider a standard power-law path-loss model with the decay rate $ \left \| d \right \|^{-\alpha } $, where $ d $ indicates the distance between the transmitter and receiver and $ \alpha $ is used to denote the path-loss exponent \cite{elsawy2014analytical}. \textcolor{black}{Given the different size and deployment locations of MBS and SBS, the path loss exponents of vehicle-SBS and that of vehicle-MBS may be different, which are denoted by  $ \alpha_{S} $, $ \alpha_{M} $, respectively.} \textcolor{black}{Furthermore, for the effects of shadowing, we need to consider the links VST, VSO, and VM. Therefore, random variables $ {\chi _{S,0}} $, $ {\chi _{S,1}} $, and $ {\chi _{M}} $ following a log-normal distribution given by $ 10log_{10}\chi_{i}\sim \aleph \left ( \omega _{i},\delta _{i}^{2} \right ) $ \cite{abdulqader2015performance}, are used to denote the shadowing effect of VST, VSO, and VM links, respectively.}
Hence, the received signal power of the typical vehicle from SBSs on the typical road, SBSs on the other roads and MBSs in DL is \cite{chetlur2019coverage}
\textcolor{black}{\begin{equation}
 P_{r,V} = \left\{{\begin{array}{*{20}{c}}
 	{{P_M}{G_{M}}{H_{M}}{\chi _{M}}{{\left\| x \right\|}^{ - {\alpha _M}}},}&{x \in  {\Phi _M}}\\ 
  {P_S}{G_{S,0}}{H_{S,0}}{\chi _{S,0}}{\left\| x \right\|^{ - {\alpha _S}}},	&x \in \Xi _{{l_0}}^{S}\\
 {P_S}{G_{S,1}}{H_{S,1}}{\chi _{S,1}}{\left\| x \right\|^{ - {\alpha _{{S}}}}} , &x \in { \Theta_{S}}\backslash \Xi _{{l_0}}^{S}.
 \end{array}}\right. 
\end{equation}}

Similarly, the received signal power in UL can be written as
\textcolor{black}{\begin{equation}
{P_{r,M}} = \begin{array}{*{20}{c}}
{{P_V}{G_{V,1}}{H_{V,1}}{\chi _M}{{\left\| x \right\|}^{ - {\alpha _M}}},}&{x \in {\Theta_{V}}}
\end{array},
\end{equation}
\begin{equation}
{P_{r,S}} = \left\{ {\begin{array}{*{20}{c}}
	{{P_V}{G_{V,0}}{H_{V,0}}{\chi _{S,0}}{{\left\| x \right\|}^{ - {\alpha _S}}},}&{x \in \Xi _{{l_0}}^{V}}\\
	{{P_V}{G_{V,1}}{H_{V,1}}{\chi _{S,1}}{{\left\| x \right\|}^{ - {\alpha _S}}},}&{x \in {\Theta_{V}}\backslash \Xi _{{l_0}}^{V}},
	\end{array}} \right. 
\end{equation} }
where the $ P_{r,V} $, $  P_{r,M} $ and $ P_{r,S} $ are the received signal power of the typical vehicle, MBS and SBS, respectively.

\textcolor{black}{Modeling the channel with shadow fading causes the received power to not be exponentially distributed \cite{8292274}. To cope with this issue, we refer to the lemma of displacement theorem in \cite{xu2016wireless} and express it as a random displacement of the location of a typical receiver \cite{madhusudhanan2014downlink}.}
Thus, $ P_{r}(x)={P}{G_{}}{\chi}{{\left\| x \right\|}^{ - {\alpha }}}  $ can be written as $ P_{r}(y)={P}{G_{}}{{\left\| y \right\|}^{ - {\alpha }}}  $, where $ y={\chi_{}^{-\frac{1}{\alpha }}}x $, and then the transformed points form a 2D homogeneous PPP with intensity $ \lambda $ to $ E\left [ {\chi_{}^{-\frac{2}{\alpha }}}  \right ]\lambda $, and the 1D PPP's intensity is $ \lambda $ to $ E\left [ {\chi_{}^{-\frac{1}{\alpha }}}  \right ]\lambda $.
\textcolor{black}{Therefore, the transformaed ${\lambda _M} = E\left[ {\chi _{M}^{ - \frac{2}{\alpha }}} \right]\lambda_m $, ${\lambda _S} = E\left[ {\chi _{S,0}^{ - \frac{1}{\alpha }}} \right]\lambda_m$, ${\lambda _V} = E\left[ {\chi _{S,0}^{ - \frac{1}{\alpha }}} \right]\lambda_m$.}
In $ { \Theta_{S}}\backslash \Xi _{{l_0}}^{S} $ and $ {\Theta_{V}}\backslash \Xi _{{l_0}}^{V} $ , \textcolor{black}{using the Theorem 1 and Assumption 1 in \cite{chetlur2019coverage},
the vehicles and SBSs can be asymptotically converged to 2D PPPs with intensity $ {\lambda _{Sa}} = E\left[ {\chi _{S,1}^{ - \frac{2}{\alpha }}} \right] {\pi} {\lambda _l}{\lambda _s} $ and $ {\lambda _{Va}} = E\left[ {\chi _{S,1}^{ - \frac{2}{\alpha }}} \right] {\pi} {\lambda _l}{\lambda _v}  $, respectively.
We use $ \Phi_M^t $, $ \Theta_{S}^t $, $ \Xi _{{l_0}}^{S,t} $, $ \Theta_{V}^t $, $ \Xi _{{l_0}}^{V,t} $ to denote the set for MBSs, SBSs, SBSs on the typical road, vheicles, vehicles on the typical road after executing the random displacement, respectively.} Hence, the value of $ P_{r,V} $, $ P_{r,M} $, and $ P_{r,S} $ after undergoing random replacement can be obtained through similar procedures as the above steps.

\vspace{-0.5cm}

\subsection{Association Policy for BSs and Vehicles}

UL/DL decoupled access C-V2X allow vehicles to choose to access different BSs for UL and DL, separately \cite{boccardi2016decouple}.
\textcolor{black}{As shown in Fig. \ref{system-model}, the four cases of joint UL/DL association in decoupled access C-V2X networks are:}
\begin{itemize}
	\item Case 1: UL = MBS 1, DL = MBS 2 
	\item Case 2: UL = SBS, DL = MBS
	\item Case 3: UL = MBS, DL = SBS
	\item Case 4: UL = SBS 1, DL = SBS 2
\end{itemize}
However, there are only two cases for the coupled access, i.e., UL/DL = MBS and UL/DL = SBS. \textcolor{black}{It is noticed that the decoupled access's Case 1 and Case 4 are connected to two BSs instead of the same BS, while the coupled access's UL depends on DL and both the UL and DL access the same BS.}
We assume that the typical vehicle is associated with MBS/SBS which yields the maximum received power (MRP) in DL. Similarly, the MBS/SBS will serve the vehicle with the MRP in UL. Thus, the typical vehicle associates with an MBS in DL if
\begin{align} 
 {P_M}{G_{{M}}}{ {{x_M}}^{ - {\alpha _M}}} > {P_S}{G_{S,0}}{ {{x_S}}^{ - {\alpha _S}}},\label{DL power}
\end{align}
\textcolor{black}{where $ x_M$, $ x_S $ denotes the nearest distances that the corresponding MBS $ \in \Phi _M^t $ and SBS $ \in \Xi _{{l_0}}^{S,t} $ to the typical vehicle.}
Otherwise, the vehicle connects to an SBS.
Similarly, the typical vehicle associates with an MBS in UL if and only if
\begin{equation}
P_V{G_{V,1}}{{{x_M}}^{ - {\alpha _M}}} > {P_V}{G_{V,0}}{ {{x_S}}^{ - {\alpha _S}}}.\label{UL power}
\end{equation}
Otherwise, the typical vehicle associates with an SBS. We substitute Eq. \eqref{DL power} and Eq. \eqref{UL power} with $ A_{M,S} = {P_M}{G_{{M}}}/{P_S}{G_{S,0}}$, $ {B_{M,S}} = {{P_V}{G_{V,1}}}/{P_V}{G_{V,0}}$:
\begin{align}
{A_{M,S}}{ {{x_{{M}}}}^{ - {\alpha _M}}} > { {{x_{{S}}}}^{ - {\alpha _S}}},\label{assoc_A} \\
B_{M,S}{ {{x_{{M}}}}^{ - {\alpha _M}}} > {{{x_{{S}}}}^{ - {\alpha _S}}},\label{assoc_B}
\end{align}
Since the transmit power of MBS is much larger than that of the SBS and vehicle, thus, $ A_{M,S} $ is larger than $B_{M,S}$. Meanwhile, the association based on MRP is then equal to that based on the minimum distance, i.e., each vehicle will communicate with the closest BS after executing the procedure of random displacement.

\vspace{-0.1cm}

\subsection{Interference}

Different frequencies are used for UL and DL, there is no interference between DL and UL. The transmission signal of MBS and SBS is interfered by the signals transmitted from the other BSs, and the typical vehicle's interference comes from other vehicles.

Because all SBSs and MBSs use the same frequency for DL and all vehicles use another frequency for UL,
in DL \cite{chetlur2018success}, the aggregate interference of the typical vehicle is composed of the interference from the MBSs $ I_M $, the interference from the SBSs on the typical road $ I_{S,0} $ and the interference from the SBSs on other roads $ I_{S,1} $. In UL, the aggregate interference is composed of the interference from the vehicles on the typical roads $ I_{V,0} $ and the interference from the vehicles on the other roads $ I_{V,1} $.
Thus, when the typical vehicle is associated with MBS or SBS in DL, the measured signal-to-noise-and-interference-ratio (SINR) is
\begin{equation}
\textit{SINR}_{M/S,D} = \frac{{{\mathop{ P}\nolimits} \left( {{{X}}_d^*} \right)}}{{{I_M} + {I_{S,0}} + {I_{S,1}} + {\sigma_D^2}}}, \label{sinrULDL}
\end{equation}
where
\begin{align}
{I_M} &= \sum\limits_{x\in {\Phi_M}}  {{P_M}{G_M}{H_M}{\left\| x \right\|}^{ - {\alpha _M}}};\nonumber\\
{\Phi}&=\left\{\begin{matrix}
\Phi _M^t\backslash X_{d}^* ,& \text{DL = MBS, d = M}\\
\Phi _M^t,& \text{DL = SBS},\\
\end{matrix}\right.
\label{inter_iM0}
\end{align}
\begin{align}
{I_{S,0}} &= \sum\limits_{x\in {\Xi}} {{P_S}{G_{S,0}}{H_{S,0}}{\left\| x \right\|}^{ - {\alpha _S}}};\nonumber\\
{\Xi}&=\left\{\begin{matrix}
\Xi _{{l_0}}^{S,t},& \text{DL = MBS},\\
\Xi _{{l_0}}^{S,t}\backslash X_{d}^* ,& \text{DL = SBS}, d = S,0,\\
\end{matrix}\right.
\end{align}
\begin{equation}
{I_{S,1}} = \sum\limits_{{x} \in  { \Theta_{S}^t}\backslash \Xi _{{l_0}}^{S,t}} {{P_S}{G_{S,1}}{H_{S,1}}{\left\| x \right\|}^{ - {\alpha _S}}}.
\label{inter_s1}
\end{equation}
Here, the $ X_d^* $ is the distance between the typical vehicle and the associated BS. When the typical vehicle connects to the MBS or SBS, the $d = M, d = S,0 $, respectively. $ \Phi _M^t\backslash X_{M}^* $ denotes that the interference $ I_M $ is from the MBSs except for the connected MBS when $ DL=MBS $. $ \Xi _{{l_0}}^{S,t}\backslash X_{S,0}^* $ denotes that the interference $ I_{S,0} $ is from the SBSs except for the connected SBS when $ DL = SBS $.
The SINR measured in UL is
\begin{equation}
\textit{SINR}_{M/S,U} = \frac{{{\mathop{ P}\nolimits} \left( {{{X}}^*} \right)}}{{{I_{V,0}} + {I_{V,1}} + {\sigma_U^2}}},
\end{equation}
where
\begin{align}
{I_{V,0}} &= \sum\limits_{x\in \Xi _{{l_0}}^{V,t}\backslash X^*} {{P_V}{G_{V,q}}{H_{V,q}}{\left\| x \right\|}^{ - {\alpha _k}}},\nonumber\\
\{k,q\}&=\left\{\begin{matrix}
\text{DL=MBS}: k = M, q = 1\\
\text{DL=SBS}: k = S, q = 0.\\
\end{matrix}\right.
\end{align}

\begin{equation}
{I_{S,1}} = \sum\limits_{{x} \in  { \Theta_{V}^t}\backslash \Xi _{{l_0}}^{V,t}} {{P_V}{G_{V,1}}{H_{V,1}}{\left\| x \right\|}^{ - {\alpha _k}}}.
\label{inter_UL}
\end{equation}
Here, since the vehicles use directional antenna and the MBSs are not along the road, when the vehicle accesses an MBS in UL, the $\alpha_k=\alpha_M, G_{V,q}=G_{V,1}$, when the vehicle accesses a SBS in UL, the $\alpha_k=\alpha_S, G_{V,q}=G_{V,0}$.
Since the noise is much smaller than the interference and the system is interference limited, the thermal noise can be neglected for the sake of convenient analysis, therefore, $  {\sigma_U^2}= {\sigma_D^2}=0 $ \cite{sattar2019spectral}, \cite{chetlur2019coverage}.

\vspace{-0.6cm}

\section{Performance Analysis and Theoretical Results}

With the help of stochastic geometry, we first derive the association probabilities and the distributions of distance for all UL/DL joint association cases. Then, based on the above results, the SE of each UL/DL link in each case, and the CP of UL/DL for MBS/SBS are derived.

\vspace{-0.5cm}

\subsection{Association Probability} \label{association probability}

According to the null probability of 1D and 2D PPP \cite{chiu2013stochastic}, the cumulative distribution functions (CDF) of $ x_S $, $ x_M $ are
\begin{align}
\label{eq1}
{F_S}\left( {{x_S}} \right) &= 1 - \exp \left( { - 2{\lambda _S} {x_{{S}}}} \right),\\
{F_M}\left( {{x_M}} \right) &= 1 - \exp \left( { - {\lambda _M}\pi x_M^2} \right).\label{cdf_2}
\end{align}
Hence, the PDFs of $ x_S $, $ x_M $ are
\begin{align}
\label{pdf_1}{f_S}\left( {{x_S}} \right) &= 2{\lambda _S}\exp \left( { -2 {\lambda _S} {x_{{S}}}} \right),\\
\label{pdf_2}{f_M}\left( {{x_M}} \right) &= 2\pi {x_M}{\lambda _M}\exp \left( { - {\lambda _M}\pi x_M^2} \right).
\end{align}

According to the association policy given by Eq. \eqref{assoc_A} and Eq. \eqref{assoc_B}, we derive the joint association probability as follows:

1) Case 1 (UL = MBS 1, DL = MBS 2): The probability that the typical vehicle is associated to MBS both in DL and UL is
\begin{align}
\Pr \left( {{{Case }} \,1} \right)=&\Pr \left( {{{{A}}_{M,S}}X_M^{^{ - {\alpha _M}}} > X_S^{^{ - {\alpha _S}}};} \right. \nonumber\\
&\quad \left. {{B_{M,S}}X_M^{^{ - {\alpha _M}}} > {X_S}^{ - {\alpha _S}}} \right).
\end{align}
Based on the consideration of ${{{A}}_{M,S}} > {B_{M,S}}$, the above formula can be written as
\begin{align}
\Pr \left( {{{Case }} \, 1} \right)& = \Pr \left( {{B_{M,S}}X_M^{^{ - {\alpha _M}}} > X_S^{^{ - {\alpha _S}}}} \right) \nonumber\\
& = \Pr \left( {{X_{{M}}} < B_{M,S}^{\frac{1}{{{\alpha _M}}}}X_S^{\frac{{{\alpha _S}}}{{{\alpha _M}}}}} \right).
\end{align}

\begin{lemma} \label{lemma1}
	The joint association probability of Case 1 can be formulated as
	\begin{align} \label{asso_prob1}
	&\Pr \left( {{{Case}} \,1} \right) \nonumber \\
	&=1 - \int_0^\infty  {\left[ {2{\lambda _S}\exp \left( { - {\lambda _M}\pi B_{M,S}^{\frac{2}{{{\alpha _M}}}}x_S^{\frac{{2{\alpha _S}}}{{{\alpha _M}}}}- 2{\lambda _S}{x_{{S}}}} \right)} \right]} d{x_{{S}}},
	\end{align}
	When ${\alpha _S} = {\alpha _M}$, $ \Pr \left(\text{Case\,1}\right) $ in closed form can be formulated as
	\begin{equation}\label{cloesed form}
	\begin{split}
	&\Pr \left( {{{Case }} \,1} \right)= \\
	&1 - \sqrt {\frac{{\lambda _S^2}}{{{\lambda _M}B_{M,S}^{\frac{2}{\alpha }}}}} \exp \left( {\frac{{\lambda _S^2}}{{{\lambda _M}\pi B_{M,S}^{\frac{2}{\alpha }}}}} \right)\textit{erfc}\left( {\frac{{{\lambda _S}}}{{\sqrt {{\lambda _M}\pi B_{M,S}^{\frac{2}{\alpha }}} }}} \right).
	\end{split}
	\end{equation}	
\end{lemma}

\begin{IEEEproof}
The proof of joint association probability of Case 1 (UL = MBS 1, DL = MBS 2) is
\begin{align}
& \Pr \left( {{{Case }} \,1} \right) \nonumber\\
&=Pr ({B_{M,S}}X_M^{^{ - {\alpha _M}}} > X_S^{^{ - {\alpha _S}}})\nonumber\\
&= {{{{{E}}_{{X_S}}}}}\left[ {\Pr \left( {{X_{{M}}} < B_{M,S}^{\frac{1}{{{\alpha _M}}}}x_S^{\frac{{{\alpha _S}}}{{{\alpha _M}}}}\left| {{X_S}} \right.} \right)} \right]\nonumber\\
&=\int_0^\infty  {{F_M}\left( {B_{M,S}^{\frac{1}{{{\alpha _M}}}}x_S^{\frac{{{\alpha _S}}}{{{\alpha _M}}}}} \right)} {f_S}\left( {{x_{{S}}}} \right)d{x_{{S}}}\nonumber\\
& \overset{(a)}{=}1 - \int_0^\infty  {\left[ {2{\lambda _S}\exp \left( { - {\lambda _M}\pi B_{M,S}^{\frac{2}{{{\alpha _M}}}}x_S^{\frac{{2{\alpha _S}}}{{{\alpha _M}}}} - 2{\lambda _S}{x_{{S}}}} \right)} \right]} d{x_{{S}}},
\end{align}
where (a) follows from the substituting $ F_{M}\left ( \cdot  \right ) $ and $ f_{S}\left ( \cdot  \right ) $ from Eq. \eqref{cdf_2} and Eq. \eqref{pdf_1} in the previous steps.
\end{IEEEproof}

2) Case 2 (UL = SBS, DL = MBS): The probability that the typical vehicle is associated to MBS in DL and SBS in UL is
\begin{align}
\vspace{-0.12cm}
&\Pr \left({{{Case \,2}}} \right)\nonumber\\
&= \Pr \left( {{{{A}}_{M,S}}X_M^{^{ - {\alpha _M}}} > X_S^{^{ - {\alpha _S}}};{B_{M,S}}X_M^{^{ - {\alpha _M}}} < X_S^{^{ - {\alpha _S}}}} \right)\nonumber\\
&= \Pr \left( {{B_{M,S}}X_M^{^{ - {\alpha _M}}} < X_S^{^{ - {\alpha _S}}} < {{{A}}_{M,S}}X_M^{^{ - {\alpha _M}}}} \right).
\end{align}

\begin{lemma}\label{lemma2}
	The joint association probability of Case 2 can be formulated as
	\begin{equation}
	\begin{aligned} \label{asso_prob2}
	&\Pr \left( \text{{Case} \,2}  \right)\\
	&= \int_0^\infty  {\left[ {2{\lambda _S}\exp \left( { - {\lambda _M}\pi B_{M,S}^{\frac{2}{{{\alpha _M}}}}x_S^{\frac{{2{\alpha _S}}}{{{\alpha _M}}}} - 2{\lambda _S}{x_{{S}}}} \right)} \right]} d{x_{{S}}}\\
	&\quad - \int_0^\infty  {\left[ {2{\lambda _S}\exp \left( { - {\lambda _M}\pi A_{M,S}^{\frac{2}{{{\alpha _M}}}}x_S^{\frac{{2{\alpha _S}}}{{{\alpha _M}}}} - 2{\lambda _S}{x_{{S}}}} \right)} \right]} d{x_{{S}}}.
	\end{aligned}
	\end{equation}
	When ${\alpha _S} = {\alpha _M}$, $ \Pr \left(\text{Case\,2}\right) $ in closed form is:
\begin{equation} 
\begin{aligned}
&\Pr \left( {{{Case \, 2}}} \right)= \\
&\sqrt {\frac{{\lambda _S^2}}{{{\lambda _M}B_{M,S}^{\frac{2}{\alpha }}}}} \exp \left( {\frac{{\lambda _S^2}}{{{\lambda _M}\pi B_{M,S}^{\frac{2}{\alpha }}}}} \right)\textit{erfc}\left( {\frac{{{\lambda _S}}}{{\sqrt {{\lambda _M}\pi B_{M,S}^{\frac{2}{\alpha }}} }}} \right)\\
&- \sqrt {\frac{{\lambda _S^2}}{{{\lambda _M}A_{M,S}^{\frac{2}{\alpha }}}}} \exp \left( {\frac{{\lambda _S^2}}{{{\lambda _M}\pi A_{M,S}^{\frac{2}{\alpha }}}}} \right)\textit{erfc}\left( {\frac{{{\lambda _S}}}{{\sqrt {{\lambda _M}\pi A_{M,S}^{\frac{2}{\alpha }}} }}} \right).
\end{aligned}
\end{equation}
\end{lemma}
\begin{IEEEproof}
The proof of joint association probability of Case 2 (UL=SBS, DL=MBS) is
\begin{equation}
\begin{aligned}
&\Pr \left( {{{Case\, 2}}} \right)\\
&\overset{(a)}{=} \Pr \left( {{X_{{M}}} < {{A}}_{M,S}^{\frac{1}{{{\alpha _M}}}}X_S^{\frac{{{\alpha _S}}}{{{\alpha _M}}}}} \right)\\
&\quad - \Pr \left( {{X_{{M}}} < B_{M,S}^{\frac{1}{{{\alpha _M}}}}X_S^{\frac{{{\alpha _S}}}{{{\alpha _M}}}}} \right)\\
&\overset{(b)}{=}{{{E}}_{{X_S}}}\left[ {\Pr \left( {{X_{{M}}} < A_{M,S}^{\frac{1}{{{\alpha _M}}}}x_S^{\frac{{{\alpha _S}}}{{{\alpha _M}}}}\left| {{X_S}} \right.} \right)} \right]\\
&\quad - {{{E}}_{{X_S}}}\left[ {\Pr \left( {{X_{{M}}} < B_{M,S}^{\frac{1}{{{\alpha _M}}}}x_S^{\frac{{{\alpha _S}}}{{{\alpha _M}}}}\left| {{X_S}} \right.} \right)} \right]\\
&\mathop  = \limits^{} \int_0^\infty  {{F_M}\left( {A_{M,S}^{\frac{1}{{{\alpha _M}}}}x_S^{\frac{{{\alpha _S}}}{{{\alpha _M}}}}} \right)} {f_S}\left( {{x_{{S}}}} \right)d{x_{{S}}}\\
&\quad - \int_0^\infty  {{F_M}\left( {B_{M,S}^{\frac{1}{{{\alpha _M}}}}x_S^{\frac{{{\alpha _S}}}{{{\alpha _M}}}}} \right)} {f_S}\left( {{x_{{S}}}} \right)d{x_{{S}}}\\
&= \int_0^\infty  {\left[ {2{\lambda _S}\exp \left( { - {\lambda _M}\pi B_{M,S}^{\frac{2}{{{\alpha _M}}}}x_S^{\frac{{2{\alpha _S}}}{{{\alpha _M}}}} - 2{\lambda _S}{x_{{S}}}} \right)} \right]} d{x_{{S}}}\\
&\quad - \int_0^\infty  {\left[ {2{\lambda _S}\exp \left( { - {\lambda _M}\pi A_{M,S}^{\frac{2}{{{\alpha _M}}}}x_S^{\frac{{2{\alpha _S}}}{{{\alpha _M}}}} - 2{\lambda _S}{x_{{S}}}} \right)} \right]} d{x_{{S}}},
\end{aligned}
\end{equation}
where Pr(Case 2) can be converted to (a). Then in (b), the result is converted to derive the expectations for $ X_s $. The remaining derivations are similar in Lemma \ref{lemma1}. 
\end{IEEEproof}

3) Case 3 (UL = MBS, DL = SBS): The probability that the typical vehicle is associated to MBS in UL and SBS in DL is
\begin{align}\label{case3}
\Pr \left( {{{Case }} \, 3} \right)=&\Pr \left( {{{{A}}_{M,S}}X_M^{^{ - {\alpha _M}}} < X_S^{^{ - {\alpha _S}}};} \right.\nonumber\\
&\quad \left. {{B_{M,S}}X_M^{^{ - {\alpha _M}}} > {X_S}^{ - {\alpha _S}}} \right),
\end{align}
Since $ A_{M,S} $ is larger than $B_{M,S}$, there is no region that satisfies Eq. $ \eqref{case3} $. Hence, $ \Pr \left( {{{Case }} \, 3} \right)= 0  $.

4) Case 4 (UL = SBS 1, DL = SBS 2): The probability that the typical vehicle is associated to SBS both in UL and DL is:
\begin{align}\label{case4}
\Pr \left( {{{Case }} \, 4} \right)&=\Pr \left( {{{{A}}_{M,S}}X_M^{^{ - {\alpha _M}}} < X_S^{^{ - {\alpha _S}}};} \right.\nonumber\\
&\quad \left. {{B_{M,S}}X_M^{^{ - {\alpha _M}}} < {X_S}^{ - {\alpha _S}}} \right)\nonumber\\
& = \Pr \left( {{A_{M,S}}X_M^{^{ - {\alpha _M}}} < X_S^{^{ - {\alpha _S}}}} \right) \nonumber\\
& = \Pr \left( {{X_{{S}}} < A_{S,M}^{\frac{1}{{{\alpha _S}}}}X_M^{\frac{{{\alpha _M}}}{{{\alpha _S}}}}} \right).
\end{align}

\begin{lemma} \label{lemma4}
	The joint association probability of Case 4 can be formulated as
	\begin{align} \label{asso+prob4}
	&\Pr \left( {{{Case \, 4}}} \right) \nonumber\\
	&= \int_0^\infty  {\left[ {2{\lambda _S}\exp \left( { - {\lambda _M}\pi A_{M,S}^{\frac{2}{{{\alpha _M}}}}x_S^{\frac{{2{\alpha _S}}}{{{\alpha _M}}}} - 2{\lambda _S}{x_{{S}}}} \right)} \right]} d{x_{{S}}},
	\end{align}
	When ${\alpha _S} = {\alpha _M}$, $ \Pr \left(Case\,4\right) $ in closed form is
	\begin{align}
	&\Pr \left( {{{Case \, 4}}} \right)\nonumber \\
	&= \sqrt {\frac{{\lambda _S^2}}{{{\lambda _M}A_{M,S}^{\frac{2}{\alpha }}}}} \exp \left( {\frac{{\lambda _S^2}}{{{\lambda _M}\pi A_{M,S}^{\frac{2}{\alpha }}}}} \right)\textit{erfc}\left( {\frac{{{\lambda _S}}}{{\sqrt {{\lambda _M}\pi A_{M,S}^{\frac{2}{\alpha }}} }}} \right).
	\end{align}
\end{lemma}
\begin{IEEEproof}
The proof of joint association probability of Case 4 (UL = SBS 1, DL = SBS 2) is
\begin{align}
&\Pr \left( {{{Case 4}}} \right)\nonumber\\
&= \Pr \left( {X_S^{^{ - {\alpha _S}}}{{ > }}{{{A}}_{M,S}}X_M^{^{ - {\alpha _M}}}} \right)\nonumber\\
&\overset{(a)}{=}1 - \Pr ({X_M} < A_{M,S}^{\frac{1}{{{\alpha _M}}}}X_S^{\frac{{{\alpha _S}}}{{{\alpha _M}}}})\nonumber\\
&\mathop  = \limits^{(b)} \int_0^\infty  {\left[ {2{\lambda _S}\exp \left( { - {\lambda _M}\pi A_{M,S}^{\frac{2}{{{\alpha _M}}}}x_S^{\frac{{2{\alpha _S}}}{{{\alpha _M}}}} - 2{\lambda _S}{x_{{S}}}} \right)} \right]} d{x_{{S}}},
\end{align}
where because the original form is hard to derive, we change to calculate the probability of its complement in $ (a) $. $ (b) $ can be derived by following the steps as adopted in Lemma \ref{lemma1}.
\end{IEEEproof}

\subsection{Distribution of Distance to BSs}
In this subsection, we derive the distance distributions of typical vehicle to its serving BSs for all feasible cases.
\begin{lemma}\label{lemma5}
	The distance distribution of Case 1 is
	\begin{align}
	{f_{{X_M}\left| {{{Case }}\,1} \right.}} = \frac{{\exp \left( { - {\lambda _S}\pi 2B_{S,M}^{\frac{1}{{{\alpha _S}}}}{x^{\frac{{{\alpha _M}}}{{{\alpha _S}}}}}} \right){f_{{X_M}}}\left( x \right)}}{{\Pr \left( {{{Case }}\,1} \right)}}.
	\end{align}
\end{lemma}

\begin{IEEEproof}
To derive the distance distribution of Case 1, we need to compute the CCDF as
\begin{align}
&F_{_{{X_M}\left| {{{Case }}\,1} \right.}}^C\nonumber\\
&\mathop  = \Pr \left( {{X_M} > x\left| {{B_{M,S}}X_M^{^{ - {\alpha _M}}} > X_S^{^{ - {\alpha _S}}}} \right.} \right)\nonumber\\
&= \frac{{\Pr \left( {{X_M} > x;{B_{M,S}}X_M^{^{ - {\alpha _M}}} > X_S^{^{ - {\alpha _S}}}} \right)}}{{\Pr \left( {{{Case }}\,1} \right)}}\nonumber\\
&= \frac{{\Pr ({X_M} > x;{X_S} > B_{S,M}^{\frac{1}{{{\alpha _S}}}}X_M^{\frac{{{\alpha _M}}}{{{\alpha _S}}}})}}{{\Pr \left( {{{Case }}\,1} \right)}}\nonumber\\
&= \frac{{\int_x^\infty  {\left[ {\exp \left( { - {\lambda _S}\pi 2B_{S,M}^{\frac{1}{{{\alpha _S}}}}x_M^{\frac{{{\alpha _M}}}{{{\alpha _S}}}}} \right){f_{{X_M}}}\left( {{x_M}} \right)} \right]} d{x_M}}}{{\Pr \left( {{{Case }}\,1} \right)}},
\end{align}
then the CDF is $ {F_{_{{X_M}\left| {{{Case }}\,1} \right.}}} = 1 - F_{_{{X_M}\left| {{{Case }}\,1} \right.}}^C $. At last, we can achieve the PDF by differentiating CDF: ${f_{{X_M}\left| {{{Case }}\,1} \right.}} = d{F_{_{{X_M}\left| {{{Case }}\,1} \right.}}}/dx$.

Similarly, the proofs of Lemma \ref{lemma5}, Lemma \ref{lemma6}, and Lemma \ref{lemma7} can be derived by following the steps as adopted here.
\end{IEEEproof}

\begin{lemma}\label{lemma6}
	The distance distributions of Case 2 are formulated as given in Eq. \eqref{dis xm case2} and Eq. \eqref{dis xs case2}.
\begin{figure*}[!b]

	\normalsize
		\hrulefill
	\begin{equation}
	\begin{aligned}
	{f_{{X_M}\left| {{{Case\, 2}}} \right.}} = \frac{{\left[ {\exp \left( { - {\lambda _S}\pi 2{{A}}_{S,M}^{\frac{1}{{{\alpha _S}}}}{x^{\frac{{{\alpha _M}}}{{{\alpha _S}}}}}} \right) - \exp \left( { - {\lambda _S}\pi 2B_{S,M}^{\frac{1}{{{\alpha _S}}}}{x^{\frac{{{\alpha _M}}}{{{\alpha _S}}}}}} \right)} \right]{f_{{X_M}}}\left( x \right)}}{{\Pr \left( {{{Case \,2}}} \right)}},\\
	\label{dis xm case2}
	\end{aligned}
	\end{equation}
	\begin{equation}
	\begin{aligned}
	{f_{{X_{{S}}}\left| {{{Case}}{\kern 1pt} {{\,2}}} \right.}} = \frac{{\left[ {\exp \left( { - {\lambda _M}\pi B_{M,S}^{\frac{2}{{{\alpha _M}}}}{x^{\frac{{{\alpha _S}}}{{{\alpha _M}}}}}} \right) - \exp \left( { - {\lambda _M}\pi A_{M,S}^{\frac{2}{{{\alpha _M}}}}{x^{\frac{{{\alpha _S}}}{{{\alpha _M}}}}}} \right)} \right]{f_{{X_S}}}\left( x \right)}}{{\Pr \left( {{{Case}}{\mkern 1mu} {{\,2}}} \right)}}.
	\label{dis xs case2}
	\end{aligned}
	\end{equation}
	\vspace*{4pt}
\end{figure*}

\end{lemma}

\begin{IEEEproof}
	Similar to the proof of Lemma \ref{lemma5}.	
\end{IEEEproof}

\begin{lemma}\label{lemma7}
	The distance distribution of Case 4 is formulated as
	\begin{align}
	{f_{{X_{{S}}}\left| {{{Case }}4} \right.}} = \frac{{\exp \left( { - {\lambda _{{M}}}\pi {{A}}_{M,S}^{\frac{2}{{{\alpha _M}}}}{x^{\frac{{2{\alpha _S}}}{{{\alpha _M}}}}}} \right){f_{{X_S}}}\left( x \right)}}{{\Pr \left( {{{Case}}{\mkern 1mu} 4} \right)}}.
	\end{align}
\end{lemma}

\begin{IEEEproof}
	Similar to the proof of Lemma \ref{lemma5}.	
\end{IEEEproof}

\vspace{0.12cm}

\subsection{Spectral Efficiency}

SE refers to the amount of data transmitted per unit of bandwidth.
In this subsection, with results of Lemma \ref{lemma1} to \ref{lemma7}, we derive the SE of each case by
leveraging stochastic geometry \cite{andrews2011tractable}.

\textcolor{black}{
The average system SE is
\begin{equation}
\label{sys_average}
SE = \sum_{i=1}^{4} \sum_{L}^{\left \{ U,D \right \}}\tau _{{Case\, i}}^{{L}}{Pr}\left ( Case\,i \right ),
\end{equation}
where $ \tau_i $ is the average SE of Case i.}

To make the statement clearer, we first derive the SE of UL for Case 2.

\begin{theorem} 
	\label{theorem 2 UL}
	The SE of UL for Case 2 is formulated as
	\begin{align}\label{se2u}
	\tau _{2}^U = \int_0^\infty  {{f_{S\left| {{2}} \right.}}\int_0^\infty  {\Pr \left[ {{H_{V,0}} > {\beta _{S,0}}{I_{U,2}}x_S^{{\alpha _S}}} \right]} dtd{x_S}},
	\end{align}
	where $ {\beta _{S,0}} = \left({e^t} - 1\right)  / \left({P_V}{G_{V,0}}\right) $, $ {I_{U,2}} = {I_{V,0}} + {I_{V,1}} $. The $ \Pr \left[ {{H_{V,0}} > {\beta _{S,0}}{I_{U,2}}x_S^{{\alpha _S}}} \right] $ is	
	\begin{align}
	&\Pr \left[ {{H_{V,0}} > {\beta _{S,0}}{I_{U,2}}x_S^{{\alpha _S}}} \right]\nonumber\\
	&= \sum\limits_{k = 0}^{{m_S} - 1} {\frac{{{{\left( { - {m_S}{\beta _{S,0}}x_S^{{\alpha _S}}} \right)}^k}}}{{k!}}{{\left[ {\frac{{{\delta ^k}}}{{\delta {j^k}}}{\zeta _{_{{I_{U,2}}}}}\left( j \right)} \right]}_{j = {m_S}{\beta _{S,0}}x_S^{{\alpha _S}}}}},
	\end{align}
	where $ {\zeta _{{I_{U,2}}}}\left( j \right) = {\zeta _{_{{I_{V,0}}}}}\left( j \right){\zeta _{_{{I_{V,1}}}}}\left( j \right) $,
	\textcolor{black}{\begin{align}
	&{\zeta _{{I_{V,0}}}}\left( j \right)\mathop  = \mathbb{K }\left ( {\lambda _V};x_S;\infty;\frac{{{P_V}{G_{V,0}}{x^{ - {\alpha _S}}}}}{{m_S}};m_S;1 \right),
	\end{align}
	\begin{align}
	&{\zeta _{{I_{V,1}}}}\left( j \right)=\mathbb{K }\left ( \pi {\lambda _{Va}};0;\infty;\frac{{{P_V}{G_{V,1}}{x^{ - {\alpha _S}}}}}{{m_{S,1}}};m_{S,1};x \right).
	\end{align} }
\end{theorem}

\begin{IEEEproof}
	The SE of Case 2 UL is
	\begin{align}
	&\tau _{2}^U=E\left[ {\ln \left( {1 + {{S\!I\!N\!}}{{{R}}_{U,S}}} \right)} \right]\nonumber\\
	&= \int_0^\infty  {{f_{S\left| {{2}} \right.}}\int_0^\infty  {\Pr \left[ {\ln \left( {1 + \frac{{{P_V}{G_{V,0}}{H_{V,0}}x_S^{{-\alpha _S}}}}{{{I_{U,2}}}}} \right) > t} \right]} dtd{x_S}} \nonumber\\
	&\mathop {{{  =  }}}\limits^{} \int_0^\infty  {{f_{S\left| {{2}} \right.}}\int_0^\infty  {\Pr \left[ {{H_{V,0}} > \frac{{\left( {{e^t} - 1} \right){I_{U,2}}}}{{{P_V}{G_{V,0}}}}x_S^{{\alpha _S}}} \right]} dtd{x_S}}\nonumber \\
	&\mathop  = \limits^{(a)} \int_0^\infty  {{f_{S\left| {{2}} \right.}}\int_0^\infty  {\Pr \left[ {{H_{V,0}} > {\beta _{S,0}}{I_{U,2}}x_S^{{\alpha _S}}} \right]} dtd{x_S}},
	\end{align}
	where $ {\beta _{S,0}} = \left({e^t} - 1\right)  / \left({P_V}{G_{V,0}}\right) $  in  (a).
	The proof of $  \Pr \left[ {{H_{V,0}} > {\beta _{S,0}}{I_{U,2}}x_S^{{\alpha _S}}} \right] $ is
	\begin{align}
	&\Pr \left[ {{H_{V,0}} > {\beta _{S,0}}{I_{U,2}}x_S^{{\alpha _S}}} \right]\nonumber\\
	&= {E_{{I_S}}}\left\{ {\Pr \left[ {{H_{V,0}} > {\beta _{S,0}}{I_{U,2}}x_S^{{\alpha _S}}} \right]} \right\}\nonumber\\
	&\mathop  = \limits^{({{a}})} {E_{{I_S}}}\left[ {\frac{{\Gamma \left( {{m_S},{m_S}{\beta _{S,0}}{I_{U,2}}x_S^{{\alpha _S}}} \right)}}{{\Gamma \left( {{m_S}} \right)}}} \right]\nonumber\\
	&\mathop  = \limits^{(b)} {E_{{I_S}}}\left[ {\sum\limits_{k = 0}^{{m_S} - 1} {\frac{{{{\left( {{m_S}{\beta _{S,0}}{I_{U,2}}x_S^{{\alpha _S}}} \right)}^k}}}{{k!}}{e^{ - {m_S}{\beta _{S,0}}{I_S}x_S^{{\alpha _S}}}}} } \right]\nonumber\\
	&= \sum\limits_{k = 0}^{{m_S} - 1} {\frac{{{{\left( { - {m_S}{\beta _{S,0}}x_S^{{\alpha _S}}} \right)}^k}}}{{k!}}{{\left[ {\frac{{{\delta ^k}}}{{\delta {j^k}}}{\zeta _{_{{I_{U,2}}}}}\left( j \right)} \right]}_{j = {m_S}{\beta _{S,0}}x_S^{{\alpha _S}}}}},
	\end{align}
	where (a) follows from the CCDF of gamma random variable
	$ H_{V,0} $, and (b) follows from the definition of incomplete
	gamma function for integer values of $ m_S $. The aggregate interference can be divided into two independent components $ I_{V,0}, I_{V,1} $, the Laplace transform of interference can be computed as product of the Laplace transforms of the two components, thus, $ {\zeta _{{I_{U,2}}}}\left( j \right) = {\zeta _{_{{I_{V,0}}}}}\left( j \right){\zeta _{_{{I_{V,1}}}}}\left( j \right) $. Similar to the proof of Lemma 5 in \cite{chetlur2019coverage}, the Laplace transforms of $ I_{V,0} $ and $ I_{V,1} $ are
\textcolor{black}{	\begin{align}
	&{\zeta _{{I_{V,0}}}}\left( j \right)\nonumber\\
	&= {E_{{I_{V,0}}}}\left[ {\exp \left( { - j{I_{V,0}}} \right)} \right]\nonumber\\
	&= {E_{{I_{V,0}}}}\left[ {\exp \left( { - j\sum\limits_{{x_S} \in  \Xi _{{l_0}}^{V,t}\backslash [ - {x_S},{x_S}]} {{P_V}{G_{V,0}}{H_{V,0}}{x^{ - {\alpha _S}}}} } \right)} \right]\nonumber\\
	&\mathop  = \limits^{(a)} {E_{\Xi _{{l_0}}^{V,t}\backslash {{X}}^*}}{E_{{H_{V,0}}}}\left[ {\prod\limits_{{x_S} \in \Xi _{{l_0}}^{V,t}\backslash [ - {x_S},{x_S}]} {{e^{ - j{P_V}{G_{V,0}}{H_{V,0}}{x^{ - {\alpha _S}}}}}} } \right]\nonumber\\
	&\mathop  = \limits^{(b)} \exp \left[ { - 2{\lambda _V}\int_{{x_S}}^{\infty}  {\left(1 - {{\left( {1{{ + }}\frac{{j{P_V}{G_{V,0}}{x^{ - {\alpha _S}}}}}{{{m_S}}}} \right)}^{ - {m_S}}}\right)dx} } \right]\nonumber\\
	&\mathop  = \limits^{(c)}\mathbb{K }\left ( {\lambda _V};x_S;\infty;\frac{{{P_V}{G_{V,0}}{x^{ - {\alpha _S}}}}}{{m_S}};m_S;1 \right),
	\end{align} }
	where  (a) follows from the independence of  $ \Xi _{{l_0}}^{V,t} $, we convert the accumulative form into accumulative multiplication. (b) follows from the Nakagami-m fading assumption and the PGFL of a 1D PPP \cite{5226957}. The proofs above are typical steps of stochastic geometry. We define a function  $ \mathbb{K }\left ( a;b,c;d;e;f \right )  $ to simplify (b) in (c), where $ a $ is the term before the integral except the constant, $ b $ is the lower limit of integration and the upper limit is  $ c $, $ d $ is the fractional term in the integral except $ j $, $ e $ is the index term, $ f\in\{x,1\} $ is the last integral term ($ x $ if there is, or 1 if not).   
\textcolor{black}{	\begin{align}
	&{\zeta _{{I_{V,1}}}}\left( j \right)\nonumber\\
	&= {E_{{I_{V,1}}}}\left[ {\exp \left( { - j{I_{V,1}}} \right)} \right]\nonumber\\
	&= {E_{{I_{V,1}}}}\left[ {\exp \left( { - j\sum\limits_{x \in {\Phi _V^t}\backslash \Xi _{{l_0}}^{V,t}} {{P_V}{G_{V,1}}{H_{V,1}}{x^{ - {\alpha _S}}}} } \right)} \right]\nonumber\\
	&{{ = }}{E_{{\Theta _S^t}\backslash \Xi _{{l_0}}^{V,t}}}{E_{{H_{V,1}}}}\left[ {\prod\limits_{x \in {\Theta _V^t}\backslash \Xi _{{l_0}}^{V,t}} {{e^{ - j{P_V}{G_{V,1}}{H_{S,1}}{x^{ - {\alpha _S}}}}}} } \right]\nonumber\\
	&\mathop  = \limits^{(a)} \exp \left[ { - 2\pi {\lambda _{Va}}\int_0^\infty  {1 - {{\left( {1{{ + }}\frac{{j{P_V}{G_{V,1}}{x^{ - {\alpha _S}}}}}{{{m_{S,1}}}}} \right)}^{ - {m_{S,1}}}}xdx} } \right]\nonumber\\
	&=\mathbb{K }\left ( \pi {\lambda _{Va}};0;\infty;\frac{{{P_V}{G_{V,1}}{x^{ - {\alpha _S}}}}}{{m_{S,1}}};m_{S,1};x \right), 
	\end{align}}
	where (a) follows from the PGFL of a 2D PPP.
\end{IEEEproof}
\begin{theorem}
	The SE of DL for Case 2 is formulated as
	\begin{align}
	\tau _{2}^D = \int_0^\infty  {{f_{M\left| {{2}} \right.}}\int_0^\infty  {\Pr \left[ {{H_M} > {\beta _M}{I_{D,2}}x_M^{{\alpha _M}}} \right]} dtd{x_M}},	
	\end{align}
	where ${\beta _M} =  \left({e^t} - 1\right) / \left({P_M}{G_M}\right) $, $ {I_{D,2}} = {I_M} + {I_{S,0}} + {I_{S,1}} $. The $\Pr \left[ {{H_M} > {\beta _M}{I_{D,2}}x_M^{{\alpha _M}}} \right] $ is
	\begin{align}
	&\Pr \left[ {{H_M} > {\beta _M}{I_{D,2}}x_M^{{\alpha _M}}} \right]\nonumber\\
	&= \sum\limits_{k = 0}^{{m_M} - 1} {\frac{{{{\left( { - {m_M}{\beta _M}x_M^{{\alpha _M}}} \right)}^k}}}{{k!}}{{\left[ {\frac{{{\delta ^k}}}{{\delta {j^k}}}{\zeta _{{I_{D,2}}}}\left( j \right)} \right]}_{j = {m_M}{\beta _M}x_M^{{\alpha _M}}}}},
	\end{align}
	where the $ {\zeta _{{I_{D,2}}}}\left( j \right) = {\zeta _{_{{I_M}}}}\left( j \right){\zeta _{_{{I_{S,0}}}}}\left( j \right){\zeta _{_{{I_{S,1}}}}}\left( j \right) $. The $  \Pr \left[ {{H_M} > {\beta _M}{I_{D,1}}x_M^{{\alpha _M}}} \right] $ are
\textcolor{black}{	\begin{align}
	\label{2DLM}
	&{\zeta _{{I_M}}}\left( j \right)= \mathbb{K }\left ( \pi {\lambda _{M}};x_M;\infty;\frac{{{P_M}{G_{M}}{x^{ - {\alpha _M}}}}}{{m_{M}}};m_{M};x \right),
	\end{align}	
	\begin{align}
	&{\zeta _{_{{I_{S,0}}}}}\left( j \right)=\mathbb{K }\left ( \lambda_S;x_1;x_2;\frac{{{P_S}{G_{S,0}}{x^{ - {\alpha _S}}}}}{{{m_{S,0}}}};m_{S,0};1\right),
	\end{align}
where  $x_1= A_{S,M}^{\frac{1}{{{\alpha _S}}}}x_M^{\frac{{{\alpha _M}}}{{{\alpha _S}}}}, x_2=B_{S,M}^{\frac{1}{{{\alpha _S}}}}x_M^{\frac{{{\alpha _M}}}{{{\alpha _S}}}} $,
	\begin{align}
	\label{2DLS1}
	&{\zeta _{_{{I_{S,1}}}}}\left( j \right)= \mathbb{K }\left (\pi  \lambda_{Sa};0;\infty;\frac{{{P_ S}{G_{S,1}}{x^{ - {\alpha _S}}}}}{{{m_{S,1}}}};m_{S,1};x\right).
	\end{align} }
\end{theorem}
\begin{IEEEproof}
	Follows by the same arguments as in the proof of Theorem \ref{theorem 2 UL}.		
\end{IEEEproof}

\begin{theorem}
	\label{theorem UL 1}
The SE of UL for Case 1 is formulated as
\begin{align}\label{se1u}
\tau _{ 1}^U = \int_0^\infty  {{f_{M\left| 1 \right.}}\int_0^\infty  {\Pr \left[ {{H_{V,1}} > {\beta _M}{I_{U,1}}x_M^{{\alpha _M}}} \right]} dtd{x_M}},
\end{align}
where $ {\beta _M} = \left({e^t} - 1\right)/ \left({P_V}{G_{V,1}}\right) $, $ {I_{U,1}} = {I_{V,0}} + {I_{V,1}} $.
For the convenience of writing, we use $ {f_{M\left| {{{1}}} \right.}}\left ( \cdot  \right ) $  to substitute $ {f_{{X_M}\left| {{{Case\,1}}} \right.}}\left ( \cdot  \right )$ and the following PDFs reduce to the same form. And the $ \Pr \left[ {{H_M} > {\beta _M}{I_{U,1}}x_M^{{\alpha _M}}} \right] $ in $ \tau _{ 1}^U $ is
\begin{align}
&\Pr \left[ {{H_M} > {\beta _M}{I_{U,1}}x_M^{{\alpha _M}}} \right]\nonumber\\
&= \sum\limits_{k = 0}^{{m_M} - 1} {\frac{{{{\left( { - {m_M}{\beta _M}x_M^{{\alpha _M}}} \right)}^k}}}{{k!}}{{\left[ {\frac{{{\delta ^k}}}{{\delta {j^k}}}{\zeta _{{I_{U,1}}}}\left( j \right)} \right]}_{j = {m_M}{\beta _M}x_M^{{\alpha _M}}}}},
\end{align}
where $ \zeta _{_{{I_{U,1}}}}\left( j \right) $ is the Laplace transform of $ I_{U,1}  $. From stochastic geometry, the accumulation becomes a multiplication relationship that greatly simplifies the calculation. Thus, $ \zeta _{_{{I_{U,1}}}}\left( j \right) = \zeta _{_{{I_{V,0}}}}\left( j \right) \zeta _{_{{I_{V,1}}}}\left( j \right)  $. The $ \zeta _{_{{I_{V,0}}}}\left( j \right) $ and  $ \zeta _{_{{I_{V,1}}}}\left( j \right)  $ are
\textcolor{black}{\begin{align}
\label{1UV0}
&{\zeta _{_{{I_{V,0}}}}}\left( j \right)=
\mathbb{K }\left (  \lambda_{V};x_M;\infty;\frac{{{P_V}{G_{V,1}}{x^{ - {\alpha _M}}}}}{{{m_{M}}}};m_{M};1\right),
\end{align}
\begin{align}
\label{1UV1}
&{\zeta _{_{{I_{V,1}}}}}\left( j \right)=\mathbb{K }\left (  \pi\lambda_{Va};0;\infty;\frac{{{P_V}{G_{V,1}}{x^{ - {\alpha _M}}}}}{{{m_{M}}}};m_{M};x\right).
\end{align} }
\end{theorem}

\begin{IEEEproof}
The proof of Theorem \ref{theorem UL 1} is similar to the proof of Case 2 in Theorem \ref{theorem 2 UL}.	
\end{IEEEproof}

\begin{theorem} 
The SE of DL for Case 1 is formulated as
\begin{align}
\tau _{1}^D = \int_0^\infty  {{f_{M\left| 1 \right.}}\int_0^\infty  {\Pr \left[ {{H_M} > {\beta _M}{I_{D,1}}x_M^{{\alpha _M}}} \right]} dtd{x_M}},
\end{align}
where ${\beta _M} = \left({e^t} - 1 \right)/{P_M}{G_M}$, $ {I_{D,1}} = {I_M} + {I_{S,0}} + {I_{S,1}} $. The $  \Pr \left[ {{H_M} > {\beta _M}{I_{D,1}}x_M^{{\alpha _M}}} \right] $ are
\begin{align}
&\Pr \left[ {{H_M} > {\beta _M}{I_{D,1}}x_M^{{\alpha _M}}} \right]\nonumber\\
&= \sum\limits_{k = 0}^{{m_M} - 1} {\frac{{{{\left( { - {m_M}{\beta _M}x_M^{{\alpha _M}}} \right)}^k}}}{{k!}}{{\left[ {\frac{{{\delta ^k}}}{{\delta {j^k}}}{\zeta _{{I_{D,1}}}}\left( j \right)} \right]}_{j = {m_M}{\beta _M}x_M^{{\alpha _M}}}}},
\end{align}
where the $ \zeta _{_{{I_M}}}\left( j \right) = \zeta _{_{{I_M}}}\left( j \right)\zeta _{_{{I_{S,0}}}}\left( j \right)\zeta _{_{{I_{S,1}}}}\left( j \right) $, the $ \zeta _{_{{I_M}}}\left( j \right) $, $ \zeta _{_{{I_{S,0}}}}\left( j \right) $ and $ \zeta _{_{{I_{S,1}}}}\left( j \right) $ are
\textcolor{black}{\begin{align}
\label{1DM}
&{\zeta _{_{{I_M}}}}\left( j \right)=\mathbb{K }\left (  \pi\lambda_{M};x_M;\infty;\frac{{{P_M}{G_{M}}{x^{ - {\alpha _M}}}}}{{{m_{M}}}};m_{M};x\right),
\end{align}
\begin{align}
&{\zeta _{_{{I_{S,0}}}}}\left( j \right)=\mathbb{K }\left (  \lambda_{S};B_{M,S}^{\frac{1}{{{\alpha _S}}}}x_M^{\frac{{{\alpha _M}}}{{{\alpha _S}}}};\infty;\frac{{{P_S}{G_{S,0}}{x^{ - {\alpha_S}}}}}{{{m_{S,0}}}};m_{S,0};1\right),
\end{align}
\begin{align}
\label{1DS1}
&{\zeta _{_{{I_{S,1}}}}}\left( j \right)=\mathbb{K }\left (  \pi\lambda_{Sa};0;\infty;\frac{{{P_S}{G_{S,1}}{x^{ - {\alpha_S}}}}}{{{m_{S,1}}}};m_{S,1};x\right).
\end{align} }
\end{theorem}
\begin{IEEEproof}
	Follows by the same arguments as in the proof of Theorem \ref{theorem 2 UL}.	
\end{IEEEproof}

\begin{theorem} \label{theorem UL 4}
The SE of UL for Case 4 is formulated as
\begin{align}
\tau _{4}^U = \int_0^\infty  {{f_{S\left| 4 \right.}}\int_0^\infty  {\Pr \left[ {{H_{V,0}} > {\beta _S}{I_{U,4}}x_S^{{\alpha _S}}} \right]} dtd{x_S}},
\end{align}
where $ {\beta _S} = \left({e^t} - 1\right) / \left({P_V}G_{V,0}\right) $, $ {I_{U,4}} = {I_{V,0}} + {I_{V,1}} $. The $ \Pr \left[ {{H_S} > {\beta _S}{I_{U,4}}x_S^{{\alpha _S}}} \right] $ is
\begin{align}
&\Pr \left[ {{H_S} > {\beta _S}{I_{U,4}}x_S^{{\alpha _S}}} \right]\nonumber\\
&= \sum\limits_{k = 0}^{{m_S} - 1} {\frac{{{{\left( { - {m_S}{\beta _S}x_S^{{\alpha _S}}} \right)}^k}}}{{k!}}{{\left[ {\frac{{{\delta ^k}}}{{\delta {j^k}}}{\zeta _{{I_{U,4}}}}\left( j \right)} \right]}_{j = {m_S}{\beta _S}x_S^{{\alpha _S}}}}},
\end{align}
where $ {\zeta _{_{{I_{U,4}}}}}\left( j \right) = {\zeta _{_{{I_{V,0}}}}}\left( j \right){\zeta _{_{{I_{V,1}}}}}\left( j \right) $, the $ \zeta _{I_{V,0}}\left( j \right) $ and  $ \zeta _{I_{V,1}}\left( j \right)  $ are the same as that in UL of Case 2.
\end{theorem}
\begin{IEEEproof}
	The proof of Theorem \ref{theorem UL 4} is similar to the proof of Case 2 in Theorem \ref{theorem 2 UL}.	
\end{IEEEproof}

\begin{theorem} \label{theorem DL 4}
The SE of DL for Case 4 is formulated as
\begin{align}\label{se4d}
\tau _{4}^D = \int_0^\infty  {{f_{S\left| 4 \right.}}\int_0^\infty  {\Pr \left[ {{H_S} > {\beta _S}{I_{D,4}}x_S^{{\alpha _S}}} \right]} dtd{x_S}},
\end{align}
where $ {\beta _S} =\left({e^t} - 1\right) / \left({P_S}{G_{S,0}}\right) $, $ {I_{D,4}} = {I_M} + {I_{S,0}} + {I_{S,1}} $. The $ \Pr \left[ {{H_S} > {\beta _S}{I_{D,4}}x_S^{{\alpha _S}}} \right] $ is
\begin{align}
&\Pr \left[ {{H_S} > {\beta _S}{I_{D,4}}x_S^{{\alpha _S}}} \right]\nonumber\\
&= \sum\limits_{k = 0}^{{m_S} - 1} {\frac{{{{\left( { - {m_S}{\beta _S}x_S^{{\alpha _S}}} \right)}^k}}}{{k!}}{{\left[ {\frac{{{\delta ^k}}}{{\delta {j^k}}}{\zeta _{{I_{D,4}}}}\left( j \right)} \right]}_{j = {m_S}{\beta _S}x_S^{{\alpha _S}}}}},
\end{align}
where $ {\zeta _{_{{I_{{{D}},4}}}}}\left( j \right) = {\zeta _{_{{I_M}}}}\left( j \right){\zeta _{_{{I_{S,0}}}}}\left( j \right){\zeta _{_{{I_{S,1}}}}}\left( j \right) $. The Laplace transforms are
\textcolor{black}{\begin{align}
\label{4DM}
&{\zeta _{_{{I_M}}}}\left( j \right)= \mathbb{K }\left (  \pi\lambda_{M};A_{M,S}^{\frac{1}{{{\alpha _M}}}}x_S^{\frac{{{\alpha _S}}}{{{\alpha _M}}}};\infty;\frac{{{P_M}{G_{M}}{x^{ - {\alpha_M}}}}}{{{m_{M}}}};m_{M};x\right),
\end{align}
\begin{align}
&{\zeta _{{I_{S,0}}}}\left( j \right)= \mathbb{K }\left ( \lambda_{S};x_S;\infty;\frac{{{P_S}{G_{S,0}}{x^{ - {\alpha_S}}}}}{{{m_{S,0}}}};m_{S,0};1\right),
\end{align}
\begin{align}
\label{4DS1}
&{{\cal L}_{{I_{S,1}}}}\left( j \right) = \mathbb{K }\left (\pi \lambda_{Sa};0;\infty;\frac{{{P_S}{G_{S,1}}{x^{ - {\alpha_S}}}}}{{{m_{S,1}}}};m_{S,1};x\right).
\end{align}  }
\end{theorem}

\begin{IEEEproof}
	The proof of Theorem \ref{theorem DL 4} is similar to the proof of Case 2 in Theorem \ref{theorem 2 UL}.	
\end{IEEEproof}

\textit{Corollary 1:} \textcolor{black}{The coupled access has two association cases as shown in Fig. \ref{system-model}} and the SE of coupled access for the two association cases is similar to the SE of UL/DL decoupled access. The proof follows the same steps as in Theorem \ref{theorem 2 UL}. Specifically, when the typical vehicle is associated with SBS, the case is equivalent to Case 4 of decoupled access \cite{sattar2019spectral}. Therefore, the SE of UL and DL in coupled access with SBS $ \tau _{S}^U $ and $ \tau _{S}^D $ are equal to $ \tau _{4}^U $ and $ \tau _{4}^D $, respectively. When the typical vehicle connects to MBS, the SE of DL is
\begin{align}
\tau _{M}^D= \int_0^\infty  {{f_{M\left|\text{MBS} \right.}}\int_0^\infty  {\Pr \left[ {{H_M} > {\beta _M}{I_{D,M}}x_M^{{\alpha _M}}} \right]} dtd{x_M}},
\end{align}
where the $ {\beta _M} = \left({e^t} - 1\right) / \left({P_M}{G_{M}}\right)$, and
\begin{align}
&\Pr \left[ {{H_M} > {\beta _M}{I_{D,M}}x_M^{{\alpha _M}}} \right]\nonumber\\
&= \sum\limits_{k = 0}^{{m_M} - 1} {\frac{{{{\left( { - {m_M}{\beta _M}x_M^{{\alpha _M}}} \right)}^k}}}{{k!}}{{\left[ {\frac{{{\delta ^k}}}{{\delta {j^k}}}{\zeta _{{I_{D,M}}}}\left( j \right)} \right]}_{j = {m_M}{\beta _M}x_M^{{\alpha _M}}}}},
\end{align}
\begin{align}
{f_{{X_M}\left| {{{MBS}}} \right.}} = \frac{{\exp \left( { - {\lambda _S}\pi 2{{A}}_{S,M}^{\frac{1}{{{\alpha _S}}}}{x^{\frac{{{\alpha _M}}}{{{\alpha _S}}}}}} \right){f_{{X_M}}}\left( x \right)}}{{\Pr \left( {M} \right)}},
\end{align}
where
\begin{align}
\Pr \left( {{{M}}} \right)\nonumber&= \Pr \left( {{A_{M,S}}X_M^{^{ - {\alpha _M}}} > X_S^{^{ - {\alpha _S}}}} \right) \nonumber\\
&=1- \Pr \left( {{{Case }} \, 4}\right),
\end{align}
where  $ \Pr \left( {{{Case }} \, 4}\right) $ is derived in Lemma \ref{lemma4}. The $ {\zeta _{{I_{D,M}}}}\left( j \right) = {\zeta _{_{{I_M}}}}\left( j \right){\zeta _{_{{I_{S,0}}}}}\left( j \right){\zeta _{_{{I_{S,1}}}}}\left( j \right) $ and ${\zeta _{_{{I_M}}}}\left( j \right)$, ${\zeta _{_{{I_{S,1}}}}}\left( j \right) $ are the same with Eq. \eqref{2DLM} and Eq. \eqref{2DLS1} in DL of Case 2. $ {\zeta _{_{{I_{S,0}}}}}\left( j \right) $ is
\textcolor{black}{
	\begin{align}
		&{\zeta _{_{{I_{S,0}}}}}\left( j \right)=\mathbb{K }\left (  \lambda_{S};A_{S,M}^{\frac{1}{{{\alpha _S}}}}x_M^{\frac{{{\alpha _M}}}{{{\alpha _S}}}};\infty;\frac{{{P_S}{G_{S,0}}{x^{ - {\alpha_S}}}}}{{{m_{S,0}}}};m_{S,0};1\right).
	\end{align} }

The SE of UL is
\begin{align}
&\tau _{M}^U = \int_0^\infty  {{f_{M\left| MBS \right.}}\int_0^\infty  {\Pr \left[ {{H_M} > {\beta _M}{I_{U,M}}x_M^{{\alpha _M}}} \right]} dtd{x_M}},
\end{align}
where $ {\beta _M} = \left ( {e^t} - 1 \right )/\left ( {P_V}{G_V} \right )$, and
\begin{align}
\label{eq2}
&\Pr \left[ {{H_M} > {\beta _M}{I_{U,M}}x_M^{{\alpha _M}}} \right]\nonumber\\
&= \sum\limits_{k = 0}^{{m_M} - 1} {\frac{{{{\left( { - {m_M}{\beta _M}x_M^{{\alpha _M}}} \right)}^k}}}{{k!}}{{\left[ {\frac{{{\delta ^k}}}{{\delta {j^k}}}{\zeta _{{I_{U,M}}}}\left( j \right)} \right]}_{j = {m_M}{\beta _M}x_M^{{\alpha _M}}}}}.
\end{align}
The $ \zeta _{_{{I_{U,1}}}}\left( j \right) = \zeta _{_{{I_{V,0}}}}\left( j \right) \zeta _{_{{I_{V,1}}}}\left( j \right)  $. And $ \zeta _{_{{I_{V,0}}}}\left( j \right) $, $ \zeta _{_{{I_{V,1}}}}\left( j \right)  $ are equal to Eq. \eqref{1UV0} and Eq. \eqref{1UV1} in UL of Case 1.

\subsection{Coverage Probability}

The CP is defined as the probability that SINR exceeds a predetermined threshold $ t $ \cite{ matracia2021coverage}. Based on the above results from Eq. \eqref{eq1} to Eq. \eqref{eq2}, we derive the CP of UL/DL for MBS/SBS in this subsection.

\begin{theorem} \label{Cp DL MBS}
The CP of DL for MBS is
\begin{align}
&\Pr (SIR_D^M > t)\nonumber\\
&= P(Case 1)P(SIR_D^M > t|1) + P(Case2)P(SIR_D^M > t|2) \nonumber\\
&= P(Case 1)\sum\limits_{k = 0}^{{m_M} - 1} {\int_0^\infty  {\frac{{{{\left( { - {m_M}{\beta _M}x_M^{{\alpha _M}}} \right)}^k}}}{{k!}}} } \times \nonumber\\
&{\left[ {\frac{{{\delta ^k}}}{{\delta {j^k}}}{\zeta _{{I_{D,M}}}}\left( j \right)} \right]_{j}} \times {f_{M|1}}dt + P(Case2) \times \nonumber\\
&\sum\limits_{k = 0}^{{m_M} - 1} {\int_0^\infty  {\frac{{{{\left( { - {m_M}{\beta _M}x_M^{{\alpha _M}}} \right)}^k}}}{{k!}}} } {\left[ {\frac{{{\delta ^k}}}{{\delta {j^k}}}{\zeta _{{I_{D,M}}}}\left( j \right)} \right]_{}}  {f_{M|2}}dt,
\end{align}
where $ j = {m_M}{\beta _M}x_M^{{\alpha _M}} $, $ {\beta _M} = {t}/{{{P_M}{G_M}}} $. $ {\zeta _{{I_{D,M}}}}\left( j \right) = {\zeta _{{I_M}}}\left( j \right){\zeta _{{I_{S,0}}}}\left( j \right){\zeta _{{I_{S,1}}}}\left( j \right) $ and the ${\zeta _{_{{I_{S,0}}}}}\left( j \right)$, ${\zeta _{_{{I_{S,1}}}}}\left( j \right) $ are the same with Eq. \eqref{1DM} to Eq. \eqref{1DS1} in DL of Case 1.
\end{theorem}
\begin{IEEEproof}
	The proof of Theorem \ref{Cp DL MBS} is similar to the proof of Case 2 in Theorem \ref{theorem 2 UL}.	
\end{IEEEproof}

\begin{theorem} \label{CP UL MBS} 
	The CP of UL for MBS is
\begin{align}
&\Pr (SIR_U^M > t)\nonumber\\
&= P(Case 1)P(S\!I\!R_U^M > t|1)\nonumber\\\
&= P(Case 1)\sum\limits_{k = 0}^{{m_M} - 1} {\int_0^\infty  {\frac{{{{\left( { - {m_M}{\beta _M}x_M^{{\alpha _M}}} \right)}^k}}}{{k!}}} } \nonumber\\
&{\left[ {\frac{{{\delta ^k}}}{{\delta {j^k}}}{\zeta _{{I_{U,M}}}}\left( j \right)} \right]_{j = {m_M}{\beta _M}x_M^{{\alpha _M}}}} \times {f_{M|1}}dt,
\end{align}
where $ j = {m_M}{\beta _M}x_M^{{\alpha _M}} $, $ {\beta _M} = {t}/{{{P_M}{G_M}}} $. $ {\zeta _{{I_{U,M}}}}\left( j \right) = {\zeta _{{I_{V,0}}}}\left( j \right){\zeta _{{I_{V,1}}}} $ and the ${\zeta _{_{{I_{V,0}}}}}\left( j \right)$, ${\zeta _{_{{I_{V,1}}}}}\left( j \right) $ are the same with Eq. \eqref{1UV0} to Eq. \eqref{1UV1} in UL of Case 1.
\end{theorem}
\begin{IEEEproof}
	The proof of Theorem \ref{CP UL MBS} is similar to the proof of Case 2 in Theorem \ref{theorem 2 UL}.	
\end{IEEEproof}

\begin{theorem} \label{CP DL SBS} The CP of DL for SBS is
\begin{align}
&\Pr (S\!I\!R_D^S > t)\nonumber\\
&= P(Case 4)P(S\!I\!R_D^S > t|1)\nonumber\\\
&= P(Case 4)\sum\limits_{k = 0}^{{m_S} - 1} {\int_0^\infty  {\frac{{{{\left( { - {m_S}{\beta _S}x_S^{{\alpha _S}}} \right)}^k}}}{{k!}}} }\nonumber\\
&{\left[ {\frac{{{\delta ^k}}}{{\delta {j^k}}}{\zeta _{{I_{D,S}}}}\left( j \right)} \right]_{j = {m_S}{\beta _S}x_S^{{\alpha _S}}}} \times {f_{S|4}}dt,
\end{align}
where $ {\beta _S} = {t}/{{{P_S}{G_S}}} $. $ {\zeta _{{I_{D,S}}}}\left( j \right) = {\zeta _{{I_M}}}\left( j \right){\zeta _{{I_{S,0}}}}\left( j \right){\zeta _{{I_{S,1}}}}\left( j \right) $ and ${\zeta _{_{{I_{M}}}}}\left( j \right)$,  ${\zeta _{_{{I_{S,0}}}}}\left( j \right)$, ${\zeta _{_{{I_{S,1}}}}}\left( j \right) $ are the same with Eq. \eqref{4DM} to Eq. \eqref{4DS1} in DL of Case 4.
\end{theorem}
\begin{IEEEproof}
	The proof of Theorem \ref{CP DL SBS} is similar to the proof of Case 2 in Theorem \ref{theorem 2 UL}.	
\end{IEEEproof}

\begin{theorem} \label{CP UL SBS}The CP of UL for SBS is 
\begin{align}
&\Pr (S\!I\!R_U^S > t) \nonumber\\
&= P(Case 2)P(S\!I\!R_U^S > t|2) + P(Case4)P(SIR_U^S > t|4)\nonumber\\
&= P(Case 2)\sum\limits_{k = 0}^{{m_S} - 1} {\int_0^\infty  {\frac{{{{\left( { - {m_S}{\beta _S}x_S^{{\alpha _S}}} \right)}^k}}}{{k!}}} }\nonumber \\
&{\left[ {\frac{{{\delta ^k}}}{{\delta {j^k}}}{\zeta _{{I_{U,S}}}}\left( j \right)} \right]_{j = {m_S}{\beta _S}x_S^{{\alpha_S}}}} \times {f_{S|2}}dt + P(Case4)\nonumber\\
&\sum\limits_{k = 0}^{{m_S} - 1} {\int_0^\infty  {\frac{{{{\left( { - {m_S}{\beta _S}x_S^{{\alpha _S}}} \right)}^k}}}{{k!}}} } {\left[ {\frac{{{\delta ^k}}}{{\delta {j^k}}}{\zeta _{{I_{U,S}}}}\left( j \right)} \right]} \times {f_{S|4}}dt,
\end{align}
where $ j = {m_S}{\beta _S}x_S^{{\alpha _S}} $, $ {\beta _S} = {t}/{{{P_V}{G_V}}} $. $ {\zeta _{{I_{U,S}}}}\left( j \right) = {\zeta _{{I_{V,0}}}}\left( j \right){\zeta _{{I_{V,1}}}}(j) $ and the ${\zeta _{_{{I_{V,0}}}}}\left( j \right)$, ${\zeta _{_{{I_{V,1}}}}}\left( j \right) $ are the same with Eq. \eqref{1UV0} to Eq. \eqref{1UV1} in UL of Case 4.
\end{theorem}
\begin{IEEEproof}
	The proof of Theorem \ref{CP UL SBS} is similar to the proof of Case 2 in Theorem \ref{theorem 2 UL}.	
\end{IEEEproof}

\renewcommand{\algorithmcfname}{Simulation}
\begin{algorithm} [htbp]
\color{black}	\caption{Simulation for UL/DL decoupled access C-V2X} 
	\label{alg-sim} 
	\KwIn{simulation number $n$, vehicle number $V $, SBS number $ s$, MBS number $ m $, road number $ L$,  simulation radius $ r $, threshold $ t_\tau, t_{CP} $, maximum vehicle speed $ v_{\max} $, maximum vehicle density $ \lambda_{\max} $;} 
	\KwOut{Association probability $\bf{p}\in R^{1\times 4} $, SE $ \bf{\tau} $, {CP};} 
	Initialize $P\leftarrow{\bf 0}_{n \times Vl} $, $\bf{\tau}\leftarrow{\bf 0}_{n \times Vl}$ , ${CP}\leftarrow{\bf 0}_{n \times Vl} $, $ v $; \\ 
	\For{$i=1;i \le n;i++$} 
	{ 
		Generate MBSs locations $ \bf{m} $ follow PPP, road locations $ \bf{l} $ follow PLP, SBSs locations $ \bf{s} $ and vehicles locations $ \bf{V} $ follow PPP on road $ l_i $ ;\\
		\For{$ l=1;l \le L;l++$} 
		{ 
			\For{$v=1;v \le V(l);v++$} 	
			{
				According to Eq. (\ref{DL power}) and Eq. (\ref{UL power}), select Case $\bf{p}$($ i $,$v+\sum_{i=1}^{l-1}V(i) $) =\{1, 2, 3, 4\};\\			
				Calculate SIR of UL/DL according Eq. (\ref{sinrULDL});\\
				$ \bf{\tau}$($ i $,$v+\sum_{i=1}^{l-1}V(i) $) = ln($1+ S\!I\!R $), {CP}($ i $,$v+\sum_{i=1}^{l-1}V(i) $) = $ S\!I\!R $;
			}
		}
		$p_o = \sum(p(i,:)==\{1,2,3,4\})/\sum V)$;\\
	${\tau}_o$ $= \sum$($ \bf{\tau} $($ i $, find($ \bf{\tau} $($ i,: $)$> t_\tau $))$> t_\tau $)/($ \sum V $);\\
	$ {CP}_o $$= \sum$( $ {CP}$ ($ i,: $)$>t_{CP} $)/($\sum V $);
		
	}
	Return ${p} = \sum$$ p_o $/$n  $, $\bf{\tau} = \sum$$\tau_o/n  $, ${CP} = \sum$$CP_o/n $;
\end{algorithm}

\begin{table}[htp]
	\centering
	\caption{\textcolor{black}{SYSTEM PARAMETERS}}
	\begin{tabular}{l|l }
			\toprule
		\hline
		\label{table1}
		Parameters  & Value \\
		\hline
	\textcolor{black}{	Macro BS transmit power $P_{M}$ (dBm)}& 46 \\
	\textcolor{black}{	Small BS transmit power $P_{S}$ (dBm)}& 20 \\
	\textcolor{black}{	Vehicle transmit power $P_{V}$ (dBm)}& 20 \\
	\textcolor{black}{	Pathloss exponent for MBS  $\alpha _{M}$} &4 \\
	\textcolor{black}{	Pathloss exponent for SBS $\alpha _{S}$ (NLOS) }&4 \\
		\textcolor{black}{Pathloss exponent for SBS $ \alpha _{S}$ (LOS) }&2 \\
		\textcolor{black}{Antenna Gain for MBS $G_{M}$(dBi) }&0 \\
	\textcolor{black}{	Antenna Gain for vehicle in typical line $G_{S0}$ (dBi)} &0 \\
	\textcolor{black}{	Antenna Gain for vehicle in other line $G_{S1}$ (dBi)} &-20 \\
	\textcolor{black}{	Noise power $\sigma _D^2 = \sigma _U^2$    } &0 \\	
	\textcolor{black}{	Maximum Speed $ v_{\max}$  (km/h)   }&120  \\
	\textcolor{black}{	Maximum density of vehicle on road $ \lambda_{\max}$  (nodes/km)   }&63  \\
	\textcolor{black}{	Line density  $ \lambda_{l}$  (1/km)   }&10  \\
	\textcolor{black}{	Mean of log-normal shadowing gain (dB)} &0 \\
	\textcolor{black}{	Std of shadowing gain for MBS (dB) }& 4\\
		\textcolor{black}{Std of shadowing gain for SBS on typical line (dB) }& 2\\	
		\textcolor{black}{Std of shadowing gain for SBS on other lines (dB) }& 4\\
		\textcolor{black}{	Nakagami-m  $ m_{S,0} $ (LOS/NLOS), $ m_{S,1} $, $ m_{M} $} & \{1,2\}, 1, 1\\
		\textcolor{black}{	Nakagami-m  $ m_{V,0} $ (LOS/NLOS), $ m_{V,1} $  }& \{1,2\}, 1\\	
		\hline
		\bottomrule 
	\end{tabular}
\end{table}

\section{Simulation Results}

In this section, simulation results are presented. 
We consider a circular area of a 3 kilometers (km) radius with a two-tier network.
\textcolor{black}{The simulation setup follows 3GPP TR 36.942 \cite{etsi2009lte} and the existing work \cite{smiljkovikj2015analysis}, \cite{chetlur2018coverage}, \cite{9599513}, \cite{elshaer2016downlink} and the values of parameters are listed in Table  \ref{table1}. \textcolor{black}{Over 100,000 independent Monte Carlo simulations are conducted in UL/DL decoupled access and coupled access scenarios, respectively, to validate the accuracy of the proposed analytical framework and theoretical results.} The detailed steps of the simulation are in Simulation \ref{alg-sim}.}

\begin{figure}[htp]	
	\centering
	\subfigure[LOS]{\label{pr-los} \includegraphics[width=0.85\hsize]{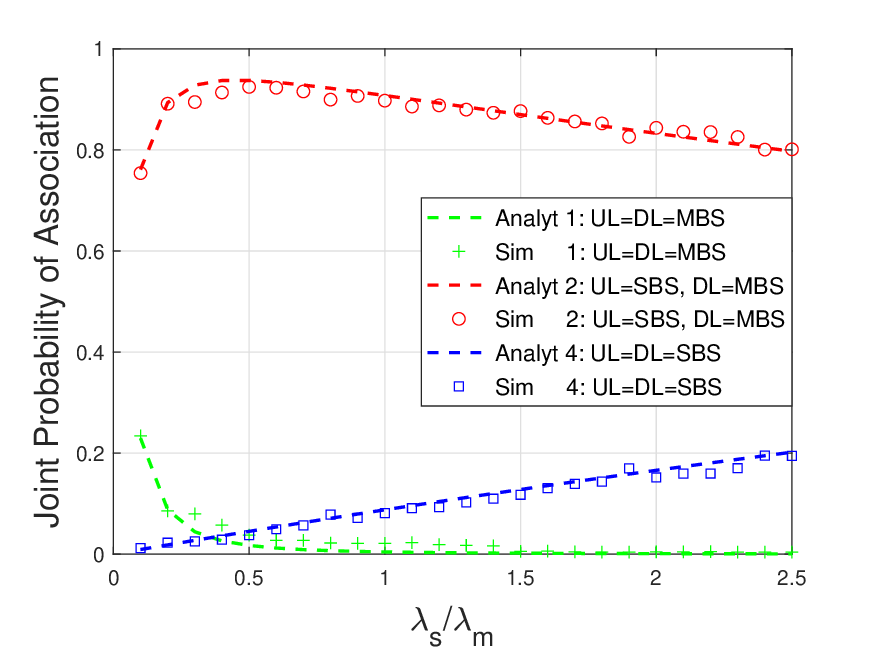}}
	
	\subfigure[NLOS]{\label{pr-nlos} \includegraphics[width=0.85\hsize]{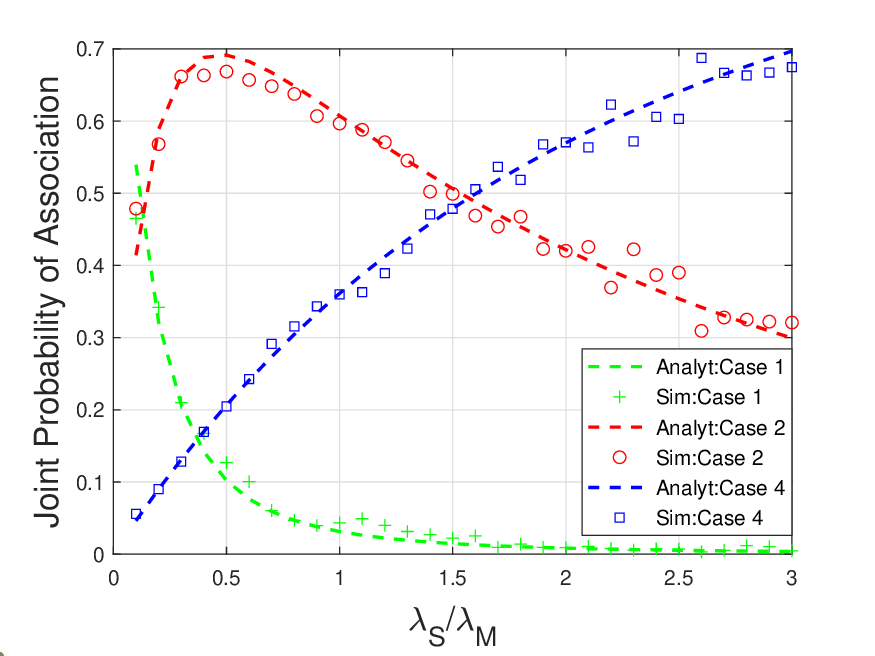}}
	
	\caption{\textcolor{black}{Joint association probability of different cases in LOS and NLOS scenarios.}}
	\label{pr}
\end{figure}

\begin{figure*}[htp]
	\subfigure[$ \lambda _{s}/\lambda _{m}=2  $, $\alpha_{S}=3$ ]{
		\label{prlambdaalpha_M2}
		\begin{minipage}[t] {0.5\linewidth}
			\centering
			\includegraphics[width=2.88in,height=1.28in]{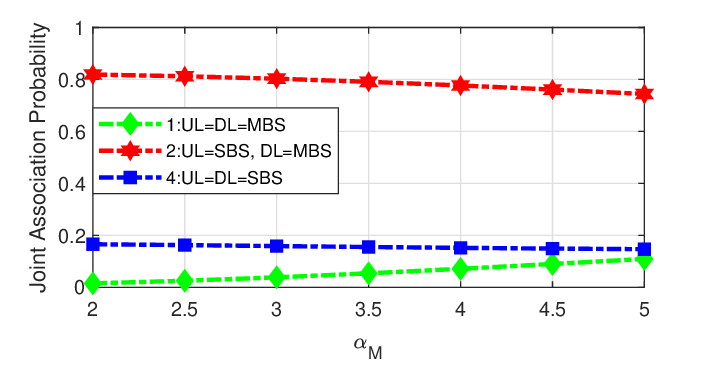}
		\end{minipage}
	}
	\subfigure[$ \lambda _{s}/\lambda _{m}=4$, $\alpha_{S}=3$]{		
		\label{prlambdaalpha_M4}
		\begin{minipage}[t] {0.5\linewidth}
			\centering
			\includegraphics[width=2.88in,height=1.28in]{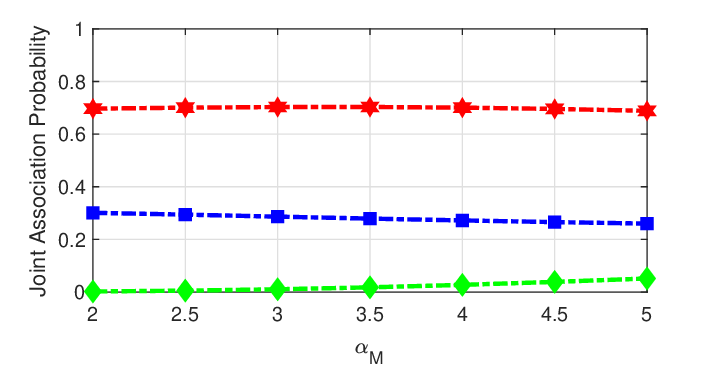}
		\end{minipage}
	}
	\subfigure[$ \lambda _{s}/\lambda _{m}=2$, $\alpha_{M}=3$]{
		\label{prlambdaalphaSM2}
		\begin{minipage}[t] {0.5\linewidth}
			\centering
			\includegraphics[width=2.88in,height=1.28in]{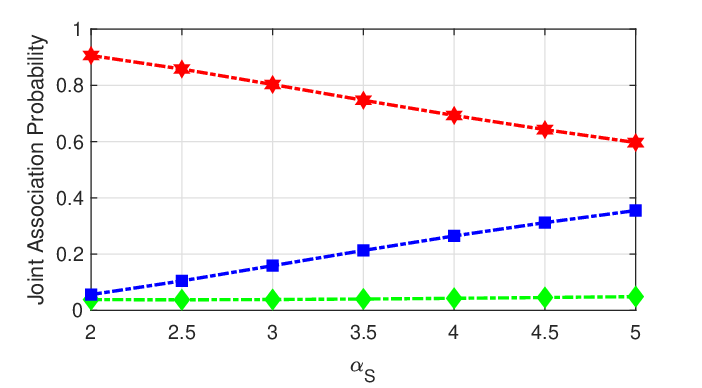}
		\end{minipage}
	}	
	\subfigure[$ \lambda _{s}/\lambda _{m}=4$, $\alpha_{M}=3$]{
		\label{prlambdaalphaSM4}
		\begin{minipage} [t] {0.5\linewidth}
			\centering
			\includegraphics[width=2.88in,height=1.28in]{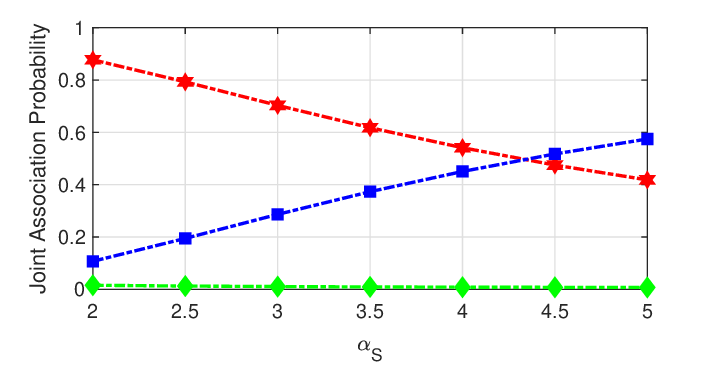}
		\end{minipage}
	}
	\caption{Joint association probability of all cases with different pathloss and different MBS/SBS densities.}
	\label{alpha-lambda}
\end{figure*}

\subsection{Joint Association Probabilities}
In Fig. \ref{pr-los} and Fig. \ref{pr-nlos}, the joint association probabilities from both analytical model and simulations for LOS and NLOS scenarios are shown. We can find that the analytical and simulation results match well, thereby demonstrating the accuracy of the proposed analytical model.
Further, more details of joint UL/DL association in decoupled C-V2X can be obtained.
In LOS scenario shown in Fig. \ref{pr-los}, the typical vehicle chooses Case 2 with a much higher probability.
This is mainly because MBS has higher transmit power than SBS in DL, while in UL, the location of SBS is closer to the typical vehicle than MBS, so SBS has smaller signal path loss and larger antenna gain when receiving from the vehicle.
In addition, since SBSs perform better than MBSs in UL, the probability of Case 1 decreases rapidly as there are more SBSs.
The probability of Case 4 is gradually increasing with SBS density, since the distance between SBS and vehicle is becoming shorter, leading to advantage of SBS's signal strength in DL against MBS.
In NLOS scenario shown in Fig. \ref{pr-nlos}, we can find the similar trend as LOS when joint association probabilities change with SBS/MBS density. An obvious difference is that the probability of Case 4 greatly increases. The main reason is that SBSs have less obstruction and path loss due to shorter range to the vehicles, thereby demonstrating the superiority of SBS in NLOS scenarios.
On the whole, decoupling can improve the load balancing of the C-V2X network as the number of SBS increases.

To better understand the characteristics of SBS and MBS, we plot the joint association probabilities against different path loss indices under different SBS/MBS densities, as shown in Fig. \ref{alpha-lambda}. Fig. \ref{prlambdaalpha_M2} and Fig. \ref{prlambdaalpha_M4} show the association probabilities under different path loss indices $\alpha_{M}$  of MBS and fixed $\alpha_{S}=3$. Fig. \ref{prlambdaalphaSM2} and Fig. \ref{prlambdaalphaSM4} show the association probabilities under different path loss indices $\alpha_{S}$  of SBS and fixed $\alpha_{M}=3$. We can see that the probability of Case 2 is decreasing as the link quality of MBS and SBS decreases (i.e., higher values of $\alpha_{}$) in the four figures of Fig. \ref{alpha-lambda}. \textcolor{black}{The joint association probabilities of the three cases are nearly constant since the MBS is not sensitive to the path loss index due to its higher transmit power as shown in Fig. \ref{prlambdaalpha_M2} and Fig. \ref{prlambdaalpha_M4}}. Also, we can see that the probability of Case 4 is increasing in Fig. \ref{prlambdaalphaSM2} and Fig. \ref{prlambdaalphaSM4}, because the $ \lambda_S $ transformed by the displacement theory decreases as $ \alpha_{S} $ increases. Therefore, after taking the transformed $ \lambda_S $ into the PDF of SBS in Eq. \eqref{pdf_1}, the PDF's value at the origin location decreases, the value at the distant point increases, and the overall function increases than before, which means the SBS's path loss index has a greater impact on vehicles close to the SBS, but it increases the association probability to vehicles far away as the $ \alpha_{S} $ increases.

{\begin{figure*}[t]	
		\centering
		\begin{minipage} [t] {0.245\textwidth} 
			\subfigure[LOS]{\label{LOS-UD-case}
				\includegraphics[width=1.1\hsize]{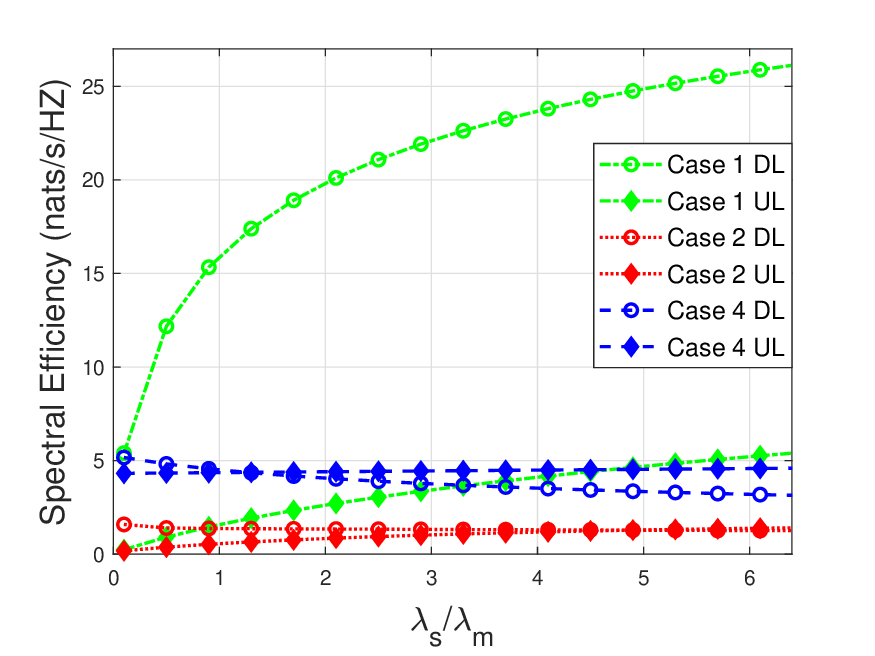}  }	 \hspace{0.04\linewidth}%
			\subfigure[NLOS]{\label{NLOS-UD-case}
				\includegraphics[width=1.1\hsize]{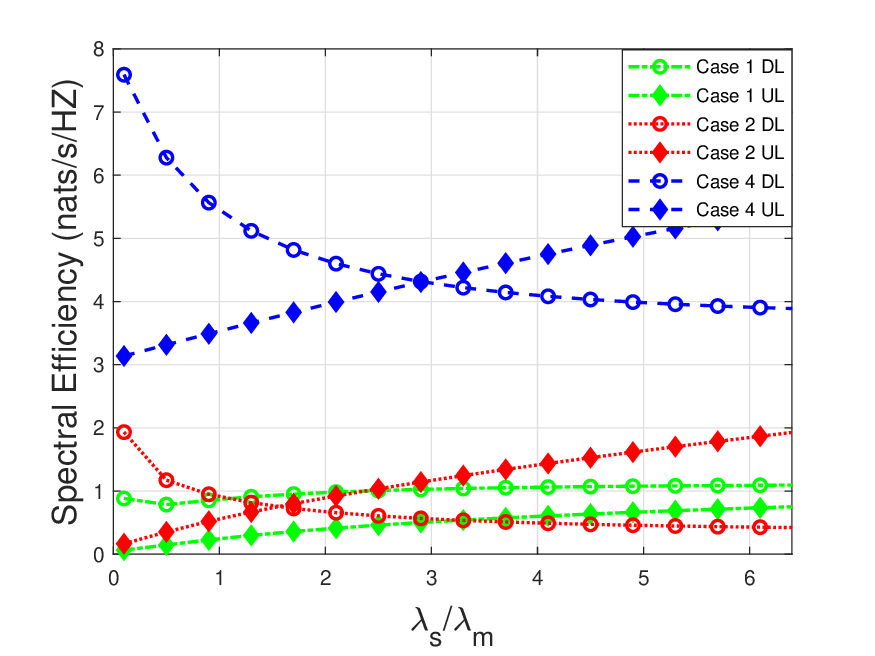} 	}	
			\caption{SE for all decoupled cases in LOS and NLOS scenarios.}
			\label{SE-UD-Case}
			\hspace{0.01\linewidth}%
		\end{minipage}
		\begin{minipage} [t] {0.245\textwidth}	
			\subfigure[LOS]{\label{LOS-se-case}
				\includegraphics[width=1.1\hsize]{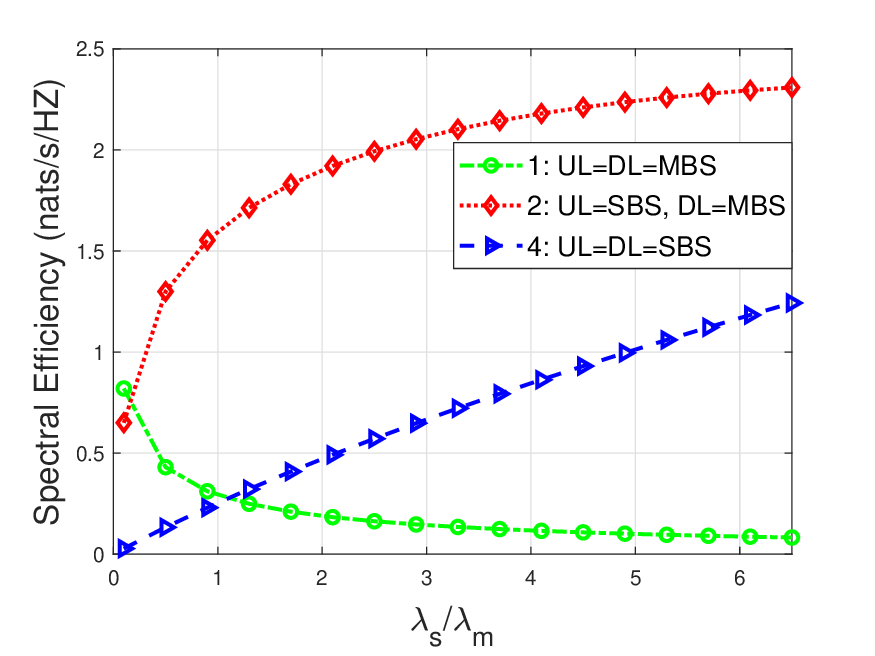}  }	 \hspace{0.04\linewidth}%
			\subfigure[NLOS]{\label{NLOS-se-case}
				\includegraphics[width=1.1\hsize]{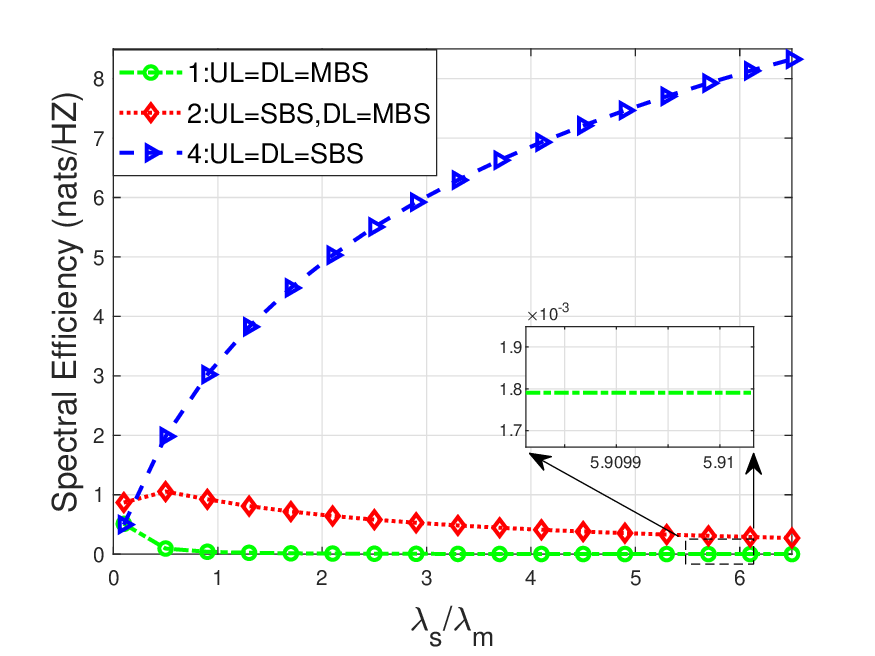} 	}	
			\caption{SE for all decoupled cases in LOS and NLOS scenarios.}
			\label{SE-Case}
			\hspace{0.01\linewidth}%
		\end{minipage}
		\begin{minipage} [t] {0.245\textwidth}	
			\subfigure[LOS]{\label{}
				\includegraphics[width=1.1\hsize]{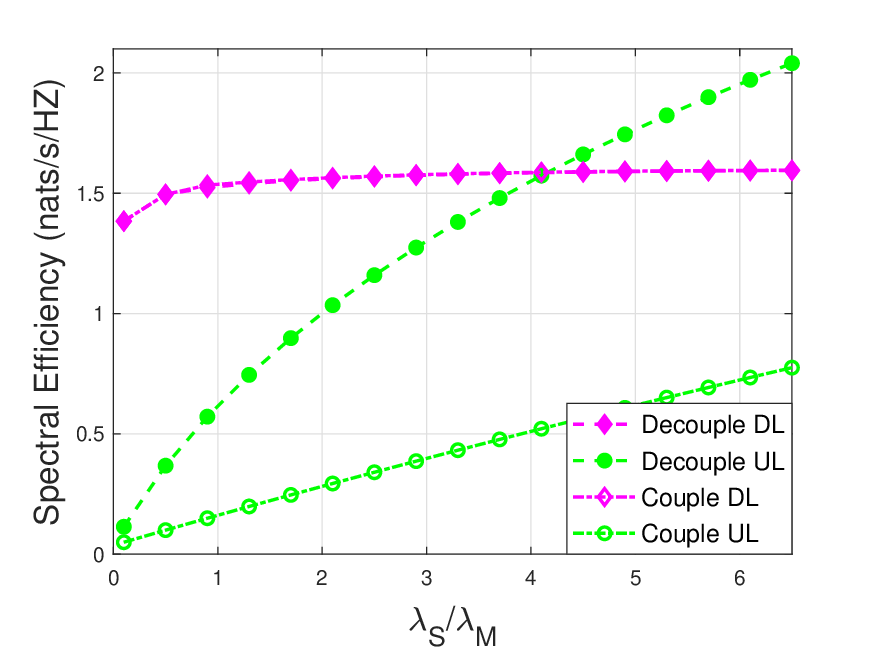}  }	 \hspace{0.04\linewidth}%
			\subfigure[NLOS]{\label{}
				\includegraphics[width=1.1\hsize]{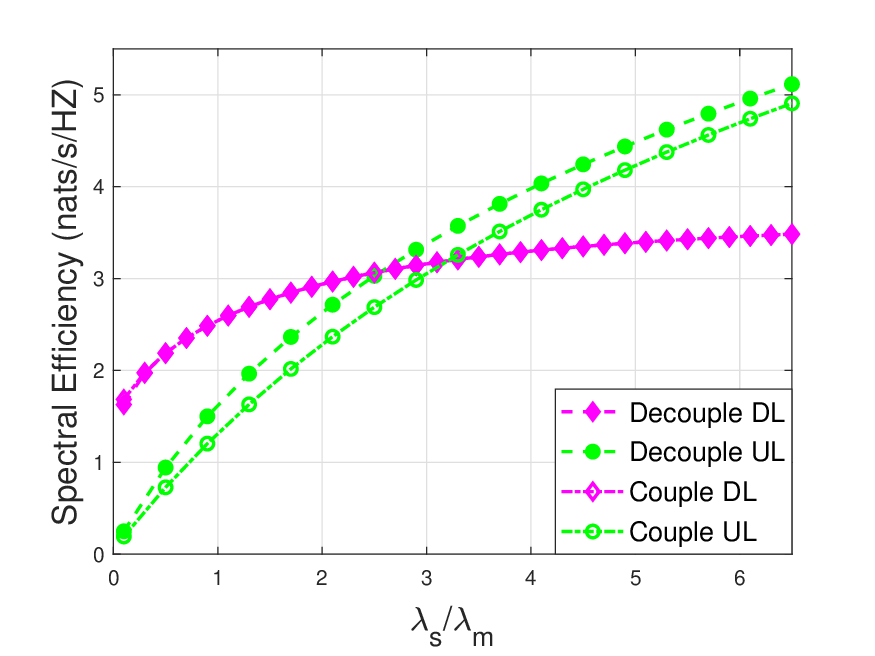} }
			\caption{SE for all decoupled and coupled links in LOS and NLOS scenarios.}
			\label{SE-DE-CO}
		\end{minipage}
		\begin{minipage} [t] {0.245\textwidth}
			\subfigure[LOS]{\label{se-los} \includegraphics[width=1.1\hsize]{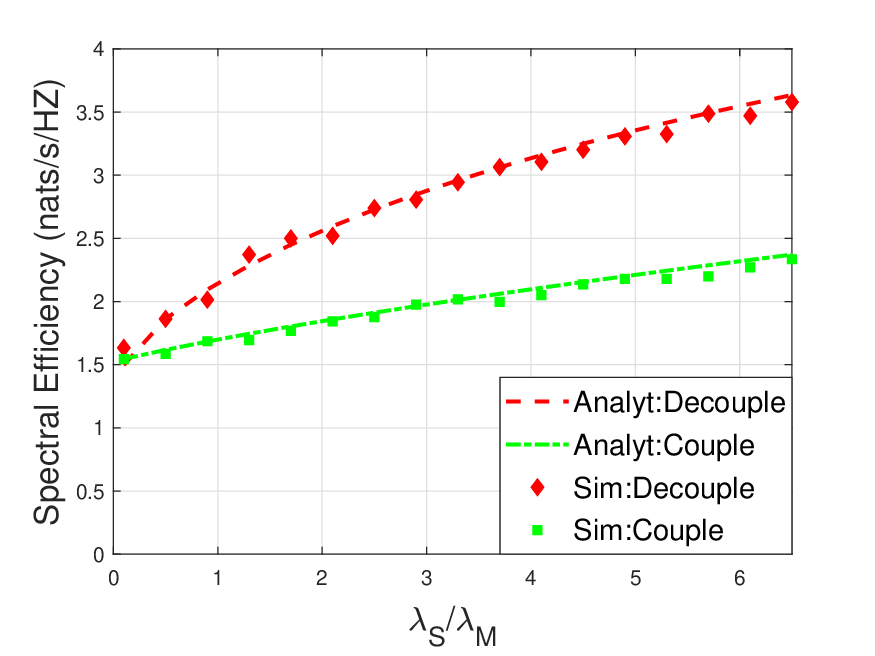}}  \hspace{0.01\linewidth}%
			\subfigure[NLOS]{\label{se-nlos} \includegraphics[width=1.1\hsize]{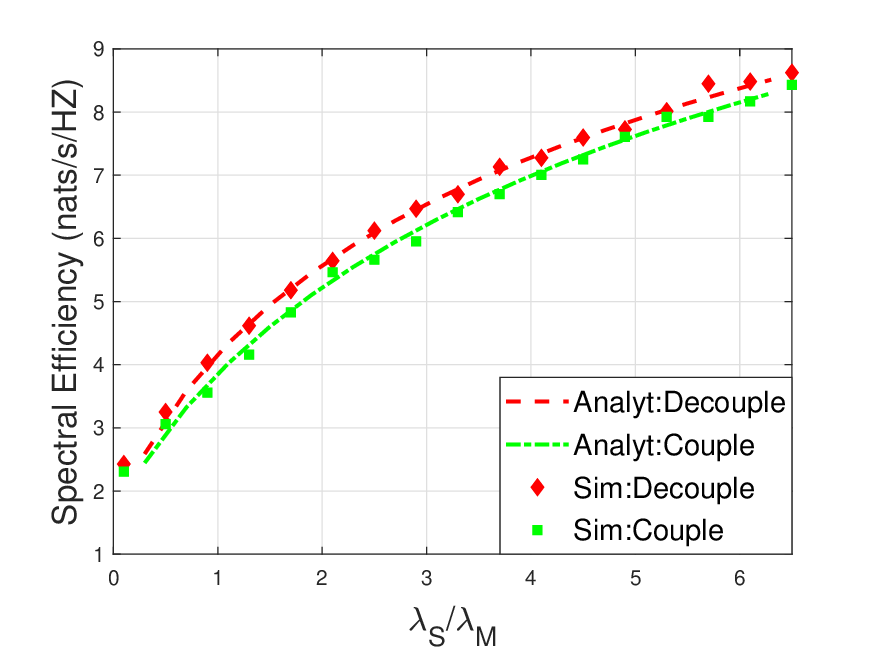}}
			\caption{System average SE for decoupled and coupled cases in LOS and NLOS scenarios.}
			\label{se}
		\end{minipage}
\end{figure*}}

\subsection{Spectral Efficiency }

In this section, we plot the SE of each link at first. Then the SE under different joint association cases is plotted and this figure can reflect the SE well while the vehicle is driving on a road. The SE of comparison between decoupled access and coupled access is plotted from 0.1 to 6.5 for the ratio of the density of SBS and MBS on the x-axis.

Fig. \ref{SE-UD-Case} shows the SE of UL/DL for all cases in LOS and NLOS. We can see that the SE of Case 1's UL and DL are increasing both in LOS and NLOS. The reason is that with the increase of SBS, the closer that the vehicles connected to the MBS, the better the SE of vehicles connected to the MBS is.
While in Fig. \ref{NLOS-UD-case}, the DL's SE of Case 2, 4 are decreasing because of the increase of SNR as the number of SBS increases.

Fig. \ref{LOS-se-case} and Fig. \ref{NLOS-se-case} show the average SE of three joint association cases in LOS and NLOS scenarios, respectively. \textcolor{black}{And it should be noted that this is not the real link SE obtained by a specific vehicle, but a weighted average SE like in Eq. \eqref{sys_average}.}
We can see that the SE of Case 1 is decreasing at the cost of increase of SE of Case 1 as $\lambda_{s} / \lambda_{m}$ increases both in LOS and NLOS. This is because of the improvement of link quality as there are more SBSs. The SE of Case 2 increases in LOS and decreases in NLOS. The reason is that most vehicles choose Case 2 in LOS and MBS still dominates SBS. While in NLOS, the reason of decreasing in SE of Case 2 is that vehicles originally connected to MBS are connected to SBS instead, resulting in increased interference in DL and the UL's SE doesn't change a lot when a vehicle is connected to MBS.

Fig. \ref{SE-DE-CO} shows the SE of UL and DL of decoupled and coupled access in LOS and NLOS scenarios. We can see that the SE is improved in UL after being decoupled over to coupled access, while the DL is the same with the SE of coupled access.
Since the DL BS is selected according to the DL reference signal power, the DL of being decoupled is the same as in coupled.
While in UL, the vehicles get rid of the limitation of UL/DL coupling through decoupling and can connect to BSs with smaller path loss and shorter distance, so as to improve the UL SE.

For further comparison between decoupled and coupled access, we plot Fig. \ref{se} to show the system average SE by using Eq. \eqref{sys_average} and the simulation results verify the effectiveness of the theoretical results.
The performance improvement shows similar trends as Fig. \ref{SE-DE-CO}.
We can see that the SE of decoupled access is nearly improved by 50\% compared with the SE of coupled access in LOS scenario. While in NLOS, the SE advantage of decoupling is not remarkable. The reason is that, in NLOS, most vehicles follow Case 4, as shown in Fig. \ref{pr-nlos}, and Case 4 is equivalent to the coupled access case in which the typical vehicle is served by the SBS, the improvement mainly comes from the UL improvement of other cases.
And we find that the decoupled mode is more beneficial for networks with higher SBS densities.

\subsection{Coverage Probability}

\begin{figure*}[t]	
	\centering
	\begin{minipage} [t] {0.32\textwidth}
		\subfigure[SBS]{\label{cov_SBS}
			\includegraphics[width=1.1\hsize]{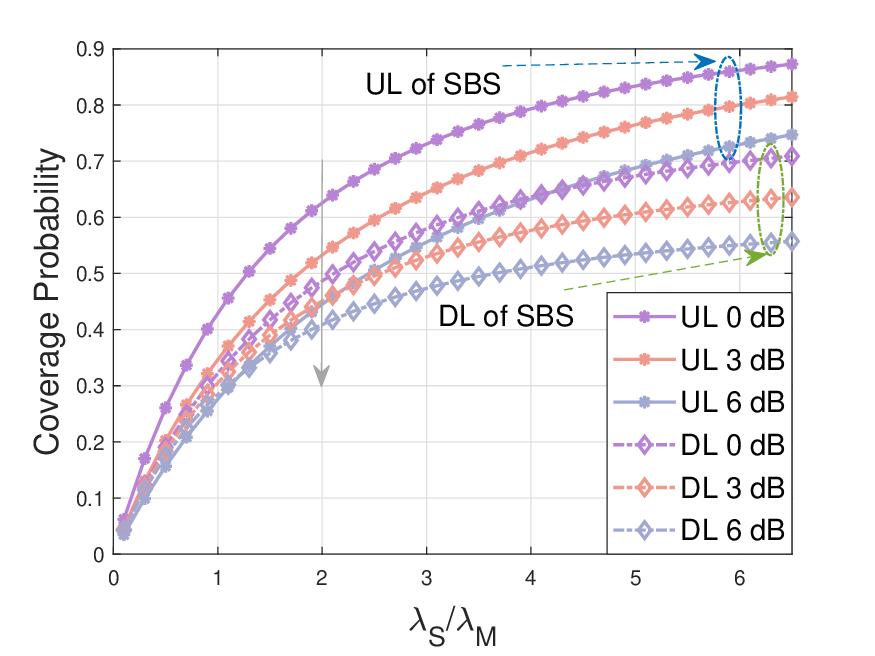}  }	 \hspace{0.04\linewidth}%
		\subfigure[MBS]{\label{cov_MBS}
			\includegraphics[width=1.1\hsize]{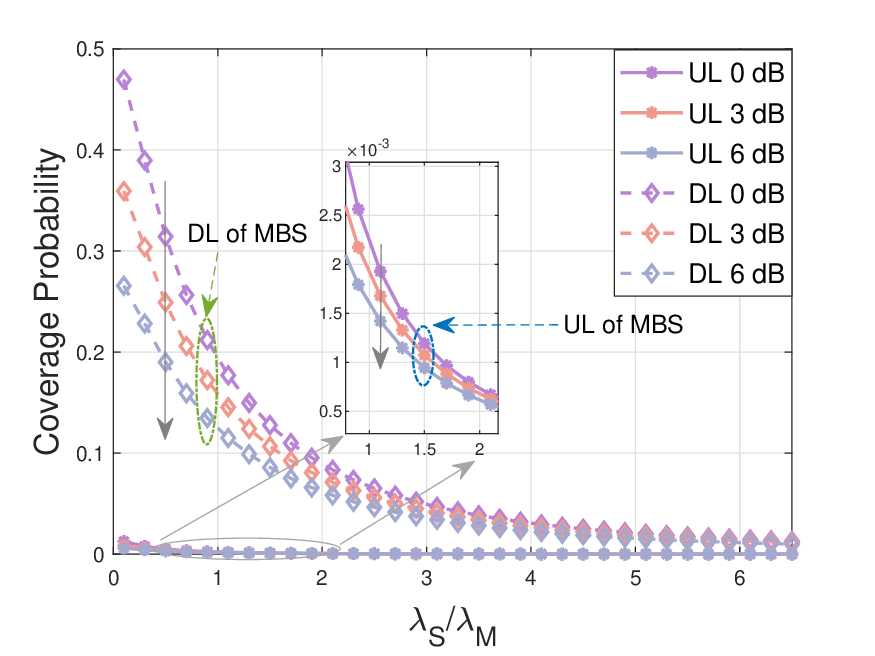} 	}	
		\caption{\textcolor{black}{CP for different thresholds ($\lambda_M = 5 $ $ nodes/km^{2} $, $\lambda_l = 1/\pi $ $ km^{-1} $, $\lambda_v = 15$ $ nodes/km $).}}
		\label{Coverage probability}
	\end{minipage}
	\begin{minipage} [t] {0.32\textwidth}
		\subfigure[SBS  ]{\label{speed cov pro}
			\includegraphics[width=1.1\hsize]{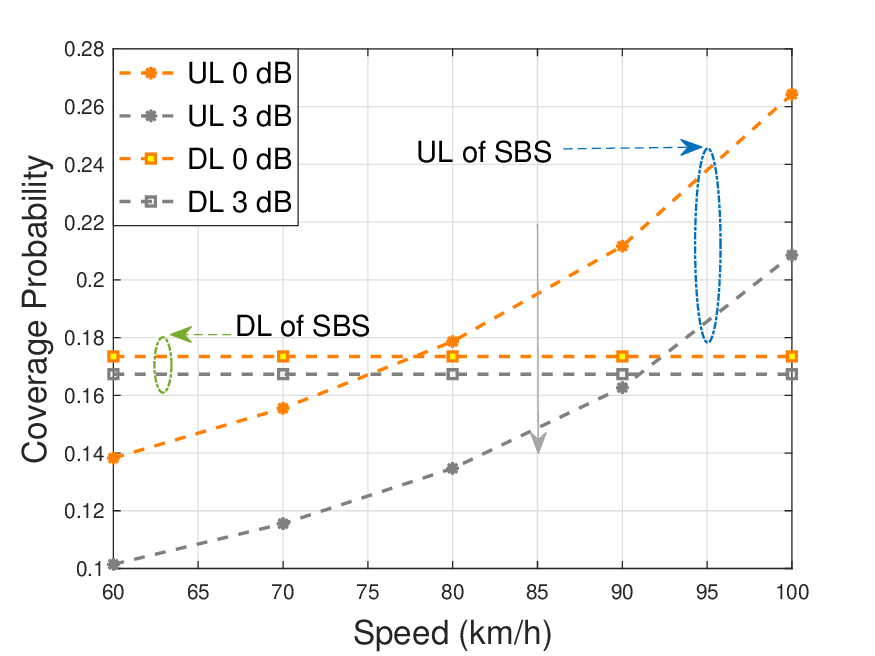}  }	 \hspace{0.04\linewidth}%
		\subfigure[MBS]{\label{speed_UD_MBS}
			\includegraphics[width=1.1\hsize]{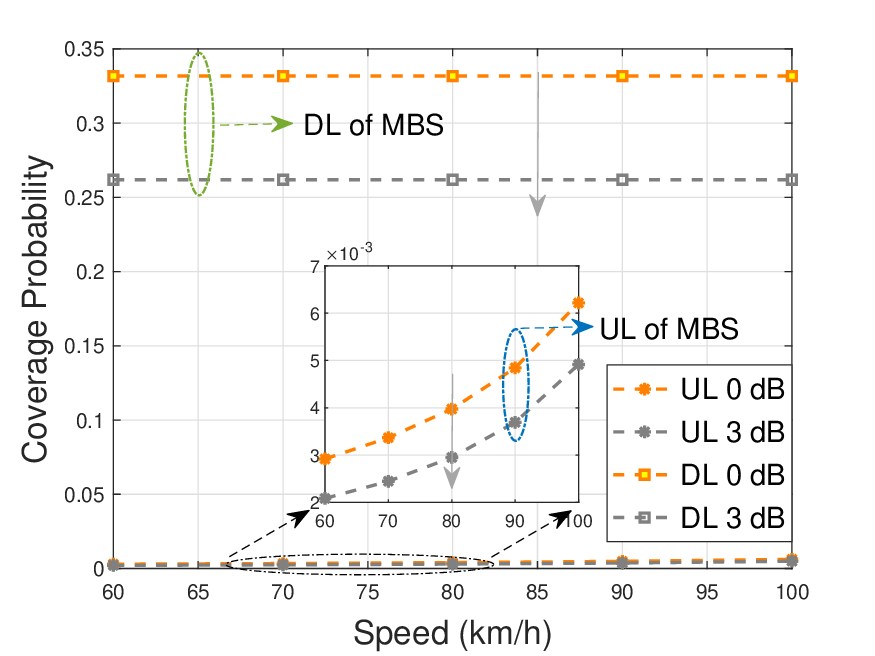} }
		\caption{ \textcolor{black}{CP of UL/DL for SBS and MBS as a function of speed ($ \lambda_s/\lambda_m = 2/4, \alpha = 4 $).}}
		\label{speed-CP}
	\end{minipage}
	\begin{minipage} [t] {0.32\textwidth}
		\subfigure[Decoupled cases]{\label{speed-se-case} 	\includegraphics[width=1.1\hsize]{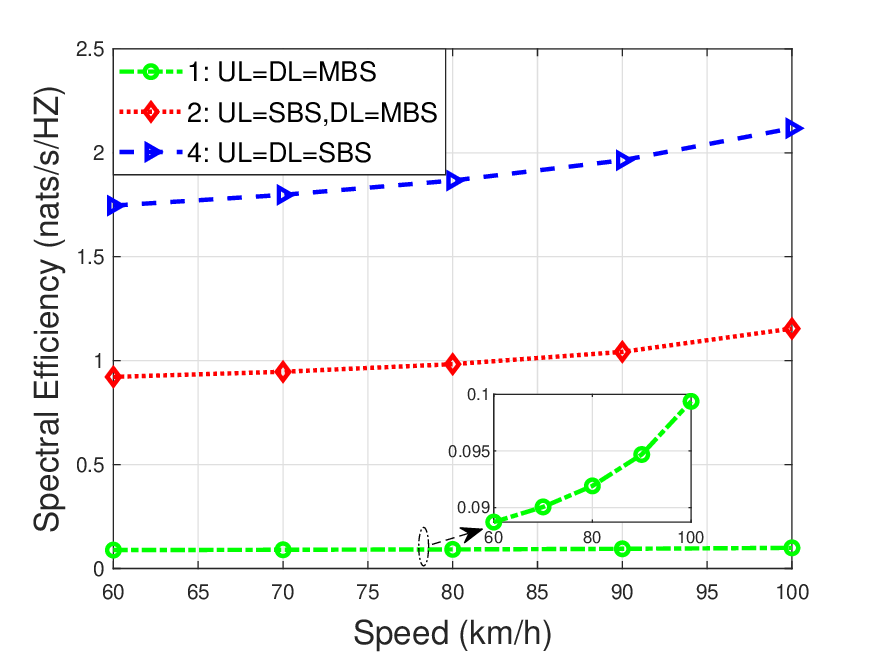}}  \hspace{0.04\linewidth}%
		\subfigure[UL/DL]{\label{speed-se-UD} 	\includegraphics[width=1.1\hsize]{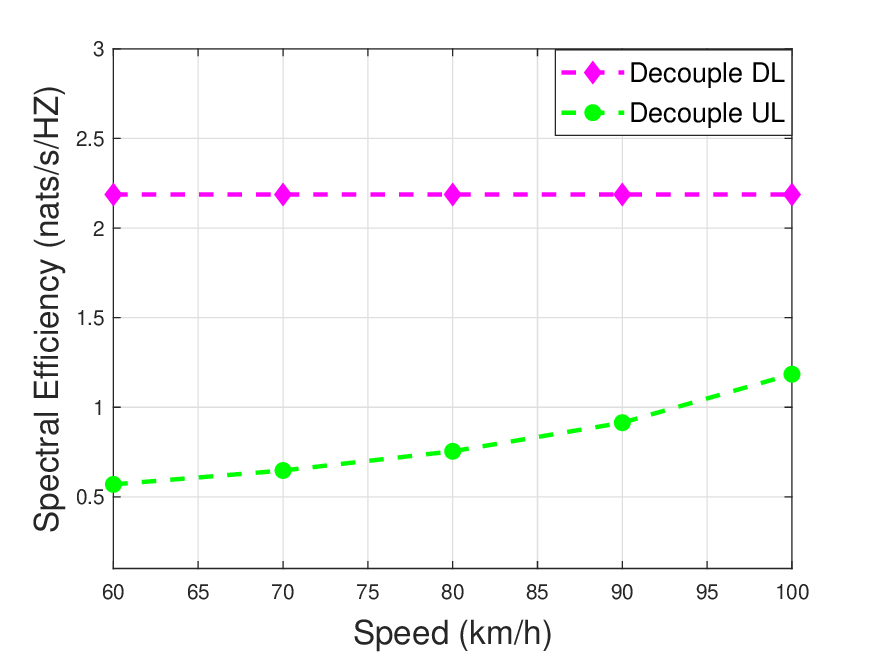}}
		\caption{\textcolor{black}{SE for different cases and UL/DL as a function of speed ($ \lambda_s/\lambda_m = 2.5/5$, $\alpha = 4$).}}
		\label{speed_se}
	\end{minipage}
\end{figure*}

Fig. \ref{Coverage probability} shows the CP of UL/DL for MBS/SBS with different SINR thresholds and different MBS/SBS densities.
Obviously, CP decreases with the increase of threshold.
It can be found that the CP of both UL and DL of SBS is increasing as $\lambda_{S} / \lambda_{M}$ increases, while the CP of both UL and DL of MBS is decreasing.
The reason is that as there are more SBSs, the distance between vehicles and SBSs becomes closer, thus the association probability of MBS decreases, leading to the decrease in the CP related with MBS (i.e., Case 1, Case 2). The reason for SBS is similar.
In a small $\lambda_{S} / \lambda_{M}$, we can find that the CP of DL for MBS is much larger than other CPs. This phenomenon also explains the reason why Case 1 and Case 2 are very large as the $\lambda_{S} / \lambda_{M}$ is small.
\vspace{0.12cm}
\subsection{The Effect of Speed}
\textcolor{black}{	
	The vehicle speed influences the vehicle density on the road \cite{liang2017resource}, and the relationship between vehicle speed $ v $ and density $\lambda$ is $v=v_{\max}\left ( 1-{\lambda }/{\lambda _{\max} }   \right )  $ according to the Greenshields Model \cite{greenshields1935study}, where $ v_{\max} $ and $ \lambda _{\max} $ denotes the maximum value of vehicle speed and density, respectively. Thus, we have $ \lambda=\lambda_{\max}\left ( 1-{v }/{v _{\max} }   \right ) $.
	In other words, the speeds of vehicles and the density of vehicles are correlated, which substantially affect the UL CP.
	}According to Eq. \eqref{asso_prob1}, \eqref{asso_prob2}, \eqref{asso+prob4}, the speed does not affect the joint association probability.
Hence, we plot the CP and SE with different speeds in Fig. \ref{speed-CP} and Fig. \ref{speed_se}, respectively.

In Fig. \ref{speed-CP}, the CP of UL is increasing, while the CP of the DL remains unchanged as the speed increases for both SBS and MBS. 
\textcolor{black}{
	This is because the speed primarily impacts the density of the vehicle, whereas the CP in DL is related with the MBS density and the effect of the vehicle speeds in DL is neglected.
	 According to the relationship between the speed and the vehicle density, a higher speed indicates lower vehicle density. Therefore, the interference in UL is reduced, leading to the increase of CP in UL.}

In Fig. \ref{speed-se-case}, we can see that the SE of different cases increases as the speed increases. The reason is that the density of vehicles is reduced as the speed increases, thereby reducing the interference in UL.
\textcolor{black}{From Fig. \ref{speed-se-UD}, we can see that the SE of DL is not affected, while the SE of UL increases as the speed increases. The reason is the same as above. Therefore, the SE gains at different speeds come from UL.}

\vspace{-0.2cm}
\section{Conclusion}

In this paper, we have done a comprehensive performance analysis of UL/DL decoupled access in C-V2X networks based on stochastic geometry.
Specifically, we have modeled the UL/DL decoupled access C-V2X as a Cox process.
We have derived the key results that characterize how a vehicle's UL/DL is associated with MBS/SBS from the statistical view, as well as the performance metrics including the joint association probability, the distance distribution of vehicle to BS, the SE, and the CP.
With the proposed analytical framework and theoretical results, an in-depth understanding of the decoupled access C-V2X network and its superiority against coupled access have been revealed. The analytical results have been further validated through extensive Monte Carlo simulations.
We hope that the work in this paper could promote decoupled access C-V2X network, such that the QoS of various C-V2X applications can be further improved.
In future work, we will focus on mobility management and dynamic access optimization in UL/DL decoupled C-V2X networks.

\vspace{-0.2cm}

\ifCLASSOPTIONcompsoc

\section*{Acknowledgments}
\else

\section*{Acknowledgment}
\fi

This work is supported in part by the National Natural Science Foundation Original Exploration Project of China under Grant 62250004, the National Natural Science
Foundation of China under Grant 62001259, 62271244, 62171217, the Innovation and Entrepreneurship of Jiangsu Province High-level Talent Program, the High-level Innovation and Entrepreneurship Talent Introduction Program Team of Jiangsu Province under Grant JSSCTD202202, the Natural Science Fund for Distinguished Young Scholars of Jiangsu Province under Grant BK20220067, and Natural Sciences and Engineering Research Council of Canada (NSERC).

\balance
\bibliographystyle{IEEEtran}
\bibliography{references}

\begin{IEEEbiography}[{\includegraphics[width=1in,height=1.25in,clip,keepaspectratio]{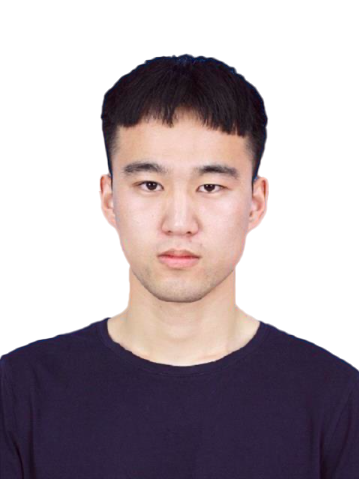}}]{Luofang Jiao} (Student Member, IEEE) received the B.S. degree in detection guidance and control technology from the University of Electronic Science and Technology of China, Chengdu, China, in 2020. He is currently working toward the Ph.D. degree with the School of Electronic Science and Engineering, Nanjing University, Nanjing, China. His research interests include uplink/downlink decoupled access, C-V2X, and heterogeneous networks.
\end{IEEEbiography}\vspace{-0.5cm}
\begin{IEEEbiography}[{\includegraphics[width=1in,height=1.25in,clip,keepaspectratio]{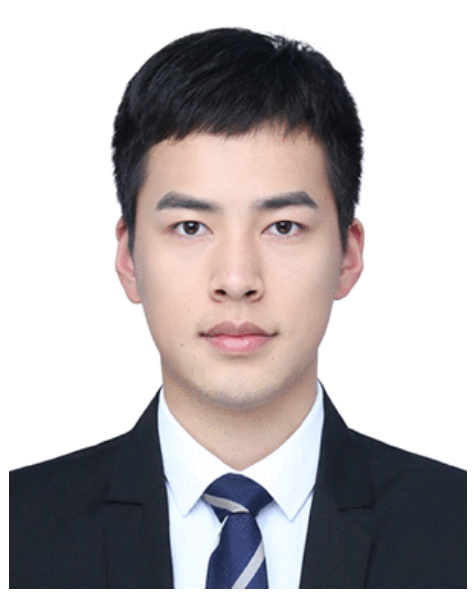}}]{Kai Yu}
	(Student Member, IEEE) received the B.S. degree in detection guidance and control technology from the University of Electronic Science and Technology of China, Chengdu, China, in 2019. He is currently working toward the Ph.D. degree with the School of Electronic Science and Engineering, Nanjing University, Nanjing, China. His research interests include resource allocation, machine learning for wireless communications, and heterogeneous networks.
\end{IEEEbiography}\vspace{-0.5cm}
\begin{IEEEbiography}[{\includegraphics[width=1in,height=1.25in,clip,keepaspectratio]{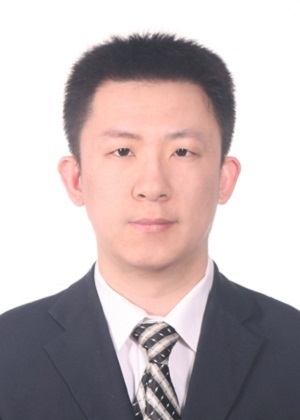}}]{Jiacheng Chen} (Member, IEEE) received the Ph.D. degree in information and communications engineering from Shanghai Jiao Tong University, Shanghai, China, in 2018. From December 2015 to December 2016, he was a Visiting Scholar with the BBCR Group, University of Waterloo, Waterloo, ON, Canada. He is currently an Assistant Researcher with Peng Cheng Laboratory, Shenzhen, China. His research interests include future network design, 5G/6G network, and resource management.
\end{IEEEbiography}\vspace{-0.5cm}
\begin{IEEEbiography}[{\includegraphics[width=1in,height=1.25in,clip,keepaspectratio]{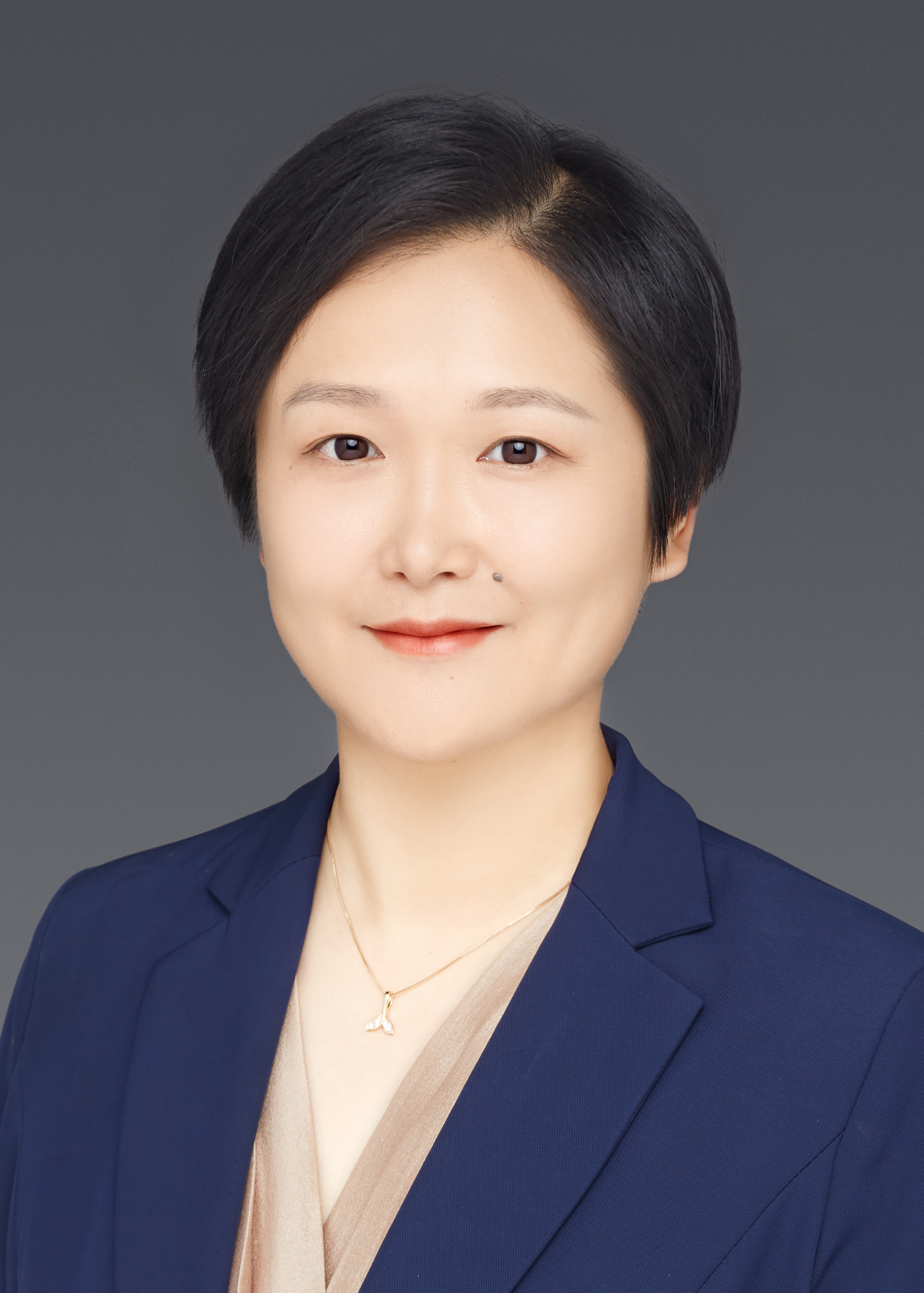}}]{Tingting Liu} (Member, IEEE) received the B.S. degree in communication engineering and the Ph.D. degree in information and communication engineering from Nanjing University of Science and Technology, Nanjing China, in 2005 and 2011, respectively. From 2017 to 2018, she was a Visiting Scholar with the University of Houston, Houston, TX, USA. She is currently a full Professor with Nanjing University of Posts and Telecommunications. Her research interests include game theory, Internet of vehicles, blockchain, caching-enabled systems, edge computing, network quality of service, device-to-device networks, and cognitive radio networks.
	
\end{IEEEbiography}\vspace{-0.5cm}
\begin{IEEEbiography}[{\includegraphics[width=1in,height=1.25in,clip,keepaspectratio]{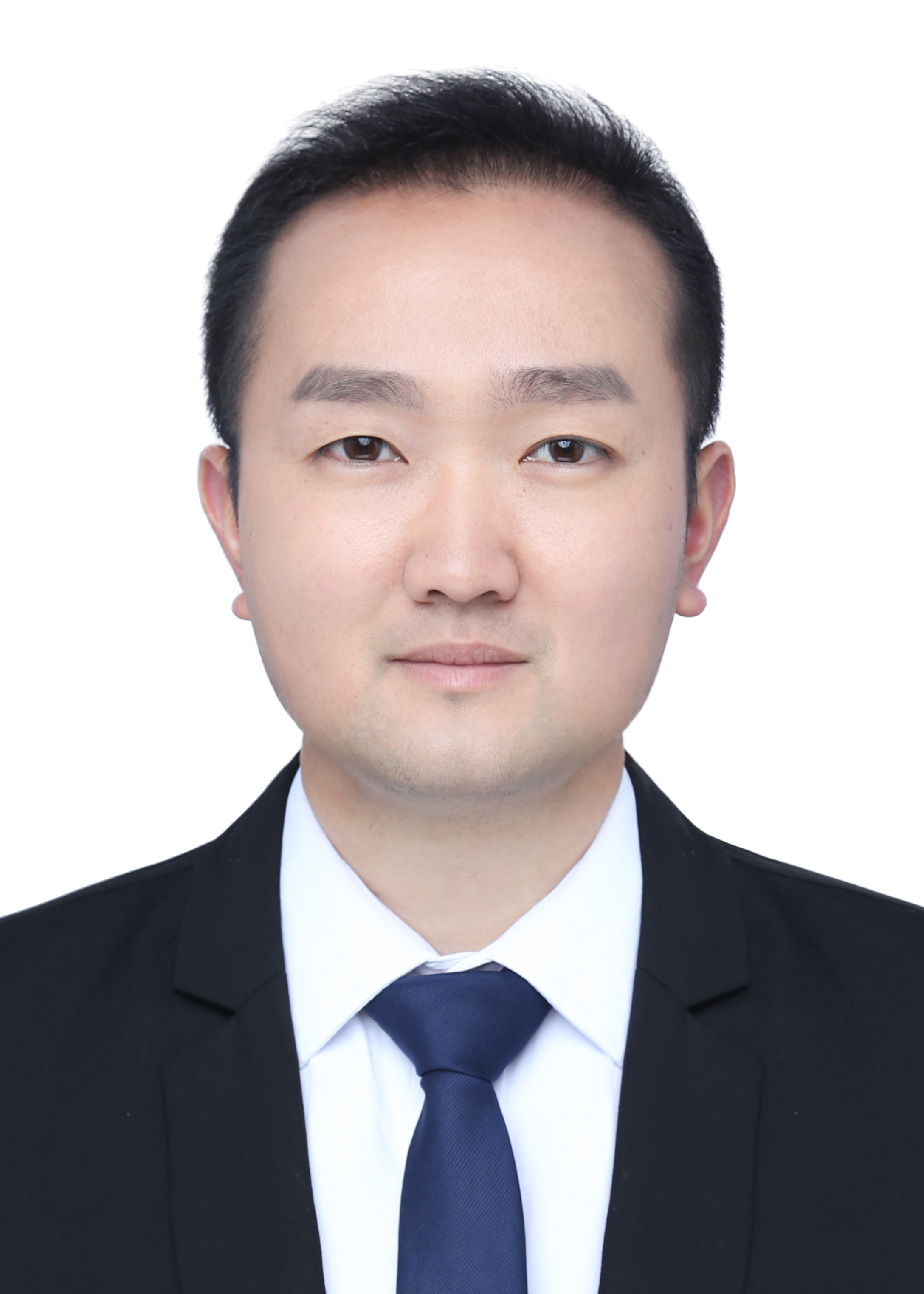}}]{Haibo Zhou} (Senior Member, IEEE) received the Ph.D. degree in information and communication engineering from Shanghai Jiao Tong University, Shanghai, China, in 2014. From 2014 to 2017, he was a Postdoctoral Fellow with the Broadband Communications Research Group, Department of Electrical and Computer Engineering, University of Waterloo. He is currently a Full Professor with the School of Electronic Science and Engineering, Nanjing University, Nanjing, China. He was elected as an IET fellow in 2022, highly cited researcher by Clarivate Analytics in 2022 \& 2020. He was a recipient of the 2019 IEEE ComSoc Asia–Pacific Outstanding Young Researcher Award, 2023-2024 IEEE ComSoc Distinguished Lecturer, and 2023-2025 IEEE VTS Distinguished Lecturer. He served as Track/Symposium CoChair for IEEE/CIC ICCC 2019, IEEE VTC-Fall 2020, IEEE VTC-Fall 2021, WCSP 2022, IEEE GLOBECOM 2022, IEEE ICC 2024. He is currently an Associate Editor of the IEEE Transactions on Wireless Communications, IEEE Internet of Things Journal, IEEE Network Magazine, and Journal of Communications and Information Networks. His research interests include resource management and protocol design in B5G/6G networks, vehicular ad hoc networks, and space-air-ground integrated networks.

\end{IEEEbiography}
\begin{IEEEbiography}[{\includegraphics[width=1in,height=1.25in,clip,keepaspectratio]{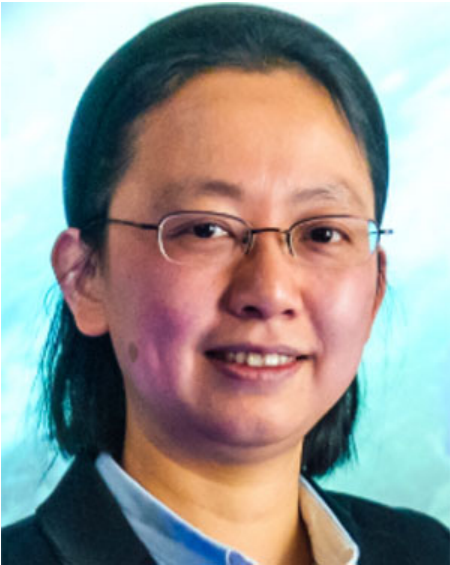}}]{Lin Cai} (Fellow, IEEE) received the M.A.Sc. and Ph.D. degrees (awarded Outstanding Achievement in Graduate Studies) in electrical and computer engineering from the University of Waterloo, Waterloo,	ON, Canada, in 2002 and 2005, respectively. Since 2005, she has been with the Department
	of Electrical and Computer Engineering, University of Victoria, Victoria, BC, Canada, and where she
	is currently a Professor. Her research interests span several areas in communications and networking,
	with a focus on network protocol and architecture design supporting emerging multimedia traffic and the Internet of Things.
	
	Prof. Cai was a recipient of the NSERC Discovery Accelerator Supplement Grants in 2010 and 2015, respectively, and the Best Paper Awards of IEEE ICC 2008 and IEEE WCNC 2011. She has co-founded and chaired the IEEE Victoria Section Vehicular Technology and Communications Joint Societies Chapter. She has been elected to serve the IEEE Vehicular Technology Society Board of Governors from 2019 to 2021. She has served as an Area Editor for the IEEE TRANSACTIONS ON VEHICULAR TECHNOLOGY, a member of the Steering Committee of the IEEE TRANSACTIONS ON BIG DATA and the IEEE TRANSACTIONS ON CLOUD COMPUTING, an Associate Editor of the IEEE INTERNET OF THINGS JOURNAL, the IEEE TRANSACTIONS ON	WIRELESS COMMUNICATIONS, the IEEE TRANSACTIONS ON VEHICULAR	TECHNOLOGY, the IEEE TRANSACTIONS ON COMMUNICATIONS, the	EURASIP Journal on Wireless Communications and Networking, the International Journal of Sensor Networks, and the Journal of Communications and Networks, and as the Distinguished Lecturer of the IEEE VTS Society. She has served as the TPC Co-Chair for IEEE VTC2020-Fall and the TPC Symposium Co-Chair for IEEE Globecom’10 and Globecom’13. She is a Registered Professional Engineer in British Columbia, Canada. She is an	NSERC E.W.R. Steacie Memorial Fellow.
\end{IEEEbiography}

\end{document}